\documentclass[11pt,onecolumn,draft]{IEEEtran}

\usepackage{epsf,psfrag,amssymb,amsfonts,cite,amsmath,yfonts}
\newtheorem{theorem}{Theorem}
\newtheorem{example}{Example}

\newtheorem{lemma}{Lemma}
\newtheorem{definition}{Definition}

\def\psfancypar#1#2{\begingroup\def\par{\endgraf\endgroup\lineskiplimit=0pt}
               \setbox2=\hbox{\large\sc #2}
               \newdimen\tmpht \tmpht \ht2 \advance\tmpht by \baselineskip
               \font\hhuge=Times-Bold at \tmpht
               \setbox1=\hbox{{\hhuge #1}}
               \count7=\tmpht \count8=\ht1
               \divide\count8 by 1000 \divide\count7 by \count8
               \tmpht=.001\tmpht\multiply\tmpht by \count7
               \font\hhuge=Times-Bold at \tmpht
               \setbox1=\hbox{{\hhuge #1}}
               \noindent
                \hangindent1.05\wd1
               \hangafter=-2 {\hskip-\hangindent
               \lower1\ht1\hbox{\raise1.0\ht2\copy1}%
                \kern-0\wd1}\copy2\lineskiplimit=-1000pt}

\newenvironment{ap}{
\renewcommand{\thesection}{\mbox{Appendix }\Alph{section}}
\renewcommand{\thesubsection}{\Alph{section}.\arabic{subsection
}}
\renewcommand{\theequation}{\mbox{\Alph{section}.}\arabic{equation}}
}{
\renewcommand{\thesection}{\arabic{section}}
\renewcommand{\thesubsection}{\thesection.\arabic{subsection}}
\renewcommand{\theequation}{\arabic{equation}}
}

\newcommand{\beq}{\begin{equation}}
\newcommand{\eeq}{\end{equation}}
\newcommand{\bqa}{\begin{eqnarray}}
\newcommand{\eqa}{\end{eqnarray}}
\newcommand{\bqn}{\begin{eqnarray*}}
\newcommand{\eqn}{\end{eqnarray*}}
\newcommand{\nn}{\nonumber}

\newcommand{\be}{\begin{enumerate}}
\newcommand{\ee}{\end{enumerate}}
\newcommand{\bi}{\begin{itemize}}
\newcommand{\ei}{\end{itemize}}
\newcommand{\bd}{\begin{description}}
\newcommand{\ed}{\end{description}}
\newcommand{\ba}{\begin{array}}
\newcommand{\ea}{\end{array}}
\newcommand{\bde}{\begin{definition}}
\newcommand{\ede}{\end{definition}}
\newcommand{\bex}{\begin{example}}
\newcommand{\eex}{\end{example}}

\def\boxit#1{\vbox{\hrule\hbox{\vrule\kern3pt
        \vbox{\kern3pt#1\kern3pt}\kern3pt\vrule}\hrule}}

\def\reals{ { {\rm  I \kern-0.15em R }  } }
\def\complex{ {\,{{\rm C} \kern-0.50em \raise0.20ex {  |}}\, }}

\def\Sigmabf{\hbox{$\bf \Sigma$}}

\def\Sigmabf{\mbox{$ \bf \Sigma $}}

\def\0bf{{\bf 0}}
\def\1bf{{\bf 1}}
\def\2bf{{\bf 2}}
\def\3bf{{\bf 3}}
\def\4bf{{\bf 4}}
\def\5bf{{\bf 5}}
\def\6bf{{\bf 6}}
\def\7bf{{\bf 7}}
\def\8bf{{\bf 8}}
\def\9bf{{\bf 9}}

\def\Abf{{\bf A}}
\def\Bbf{{\bf B}}
\def\Cbf{{\bf C}}
\def\Dbf{{\bf D}}

\def\Ibf{{\bf I}}

\def\Rbf{{\bf R}}

\def\Xbf{{\bf X}}

\def\np{{\pmb n}}
\def\up{{\pmb u}}

\def\wp{{\pmb w}}
\def\xp{{\pmb x}}
\def\yp{{\pmb y}}
\def\zp{{\pmb z}}

\def\Amat{\mathcal{A}}
\def\Bmat{\mathcal{B}}

\def\Nmat{\mathcal{N}}

\def\Rmat{\mathcal{R}}
\def\Smat{\mathcal{S}}
\def\Tmat{\mathcal{T}}

\def\QED{\mbox{\rule[0pt]{1.3ex}{1.3ex}}}
\def\bpf{{\em Proof: }}
\def\epf{\hspace*{\fill}~\QED\par\endtrivlist\unskip}

\def\Rxx{\Rbf_{\ssstyle X\kern-.1em X}}

\let\ssstyle=\scriptscriptstyle

\def\Cov{{\textrm{Cov}}}
\def\Var{{\textrm{Var}}}
\def\tr{{\textrm{tr}}}

\def\diag{{\textrm{diag}}}

\def\Kout{\setbox1=\hbox{\Huge\bf K}\hbox to
1.05\wd1{\hspace{.05\wd1}
}}
\def\Sout{\setbox1=\hbox{\Huge\bf S}\hbox to 1.05\wd1{\hspace{.05\wd1}
}}

\def\scalefig#1{\epsfxsize #1\textwidth}
\def\mi{{\underline{i}}}
\def\mj{{\underline{j}}}
\def\mk{{\underline{k}}}
\def\mr{{\underline{r}}}

\allowdisplaybreaks[2]
\begin{document}
\title{Noisy-interference Sum-rate Capacity of Parallel Gaussian Interference Channels}
\author{Xiaohu Shang, Biao Chen, Gerhard Kramer and H. Vincent Poor\thanks{X. Shang
is with Princeton University, Department of EE, Engineering Quad,
Princeton, NJ, 08544. Email: xshang@princeton.edu. B. Chen is with
Syracuse University, Department of EECS, 335 Link Hall, Syracuse,
NY 13244. Phone: (315)443-3332. Email: bichen@ecs.syr.edu. G.
Kramer was with Bell Labs, Alcatel-Lucent. He is now with
University of Southern California, Department of EE, 3740
McClintock Ave, Los Angeles, CA 90089. Email: gkramer@usc.edu. H.
V. Poor is with Princeton University, Department of EE,
Engineering Quad, Princeton, NJ, 08544. Email:
poor@princeton.edu}} \maketitle {\footnotetext{This work was
supported in part by the National Science Foundation under Grant
CNS-06-25637.}}

\begin{abstract}

The sum-rate capacity of the parallel Gaussian interference
channel is shown to be achieved by independent transmission across
sub-channels and treating interference as noise in each
sub-channel if the channel coefficients and power constraints
satisfy a certain condition. The condition requires the
interference to be weak, a situation commonly encountered in,
e.g., digital subscriber line transmission. The optimal power
allocation is characterized by using the concavity of sum-rate
capacity as a function of the power constraints.

\end{abstract}


 \maketitle

\section{Introduction}

\begin{figure}[htp]
\centerline{\begin{psfrags} \psfrag{TX1}[c]{TX1}
\psfrag{TX2}[c]{TX2} \psfrag{RX1}[c]{RX1} \psfrag{RX2}[c]{RX2}
\psfrag{vdots}[c]{\large $\vdots$} \psfrag{IFC1}[c]{IC $1$}
\psfrag{IFCm}[c]{IC $m$} \scalefig{.8}\epsfbox{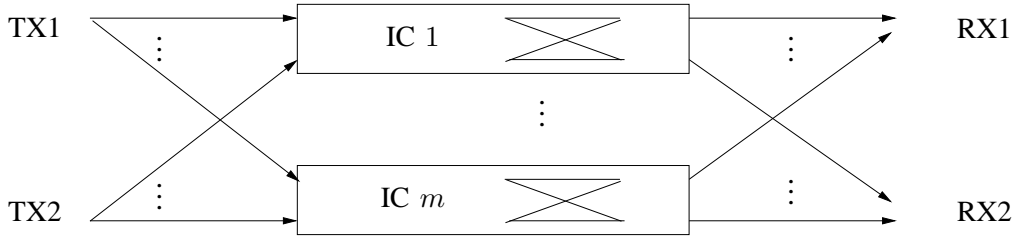}
\end{psfrags}}
\caption{\label{fig:PGIC_pifc} Illustration of a PGIC system where
the two transceiver pairs have access to $m$ independent parallel
channels.}
\end{figure}

Parallel Gaussian interference channels (PGICs) model the
situation in which several transceiver pairs communicate through a
number of independent sub-channels, with each sub-channel being a
Gaussian interference channel. Fig. \ref{fig:PGIC_pifc}
illustrates a two-user PGIC where a pair of users, each subject to
a total power constraint, has access to a set of $m$ Gaussian
interference channels (GIC). Existing systems that can be
accurately modelled as PGICs include both wired systems such as
digital subscriber lines (DSL) and  wireless systems employing
orthogonal frequency division multiple access (OFDMA). Both of
these systems have been and will be major players in broadband
systems.

While there have been extensions of information theory for the
classical single-channel GIC to the PGIC (e.g,
\cite{Chung&Cioffi:07COM}), most existing research, especially for
DSL systems, often relies on the following two assumptions
\cite{Yu-etal:02JSAC,Song-etal:02COMMAG,Cendrillon-etal:06COM}:
\bi%
\item Transmissions in sub-channels are independent of each other.
\item Each receiver treats interference as noise. %
\ei
 These assumptions greatly simplify an otherwise intractable
problem. The independent transmission assumption ensures that the
total sum rate is expressed as a sum of all sub-channels' sum
rates. The assumption of using single-user detection permits a
simple closed-form expression for the rate pair of each
sub-channel \footnote{Even under this simplified assumption,
finding the optimal power allocation is
 an NP hard problem \cite{Hayashi&Luo:ITsubmission}}. Our main
  goal in this paper is to provide a sound theoretical basis for such
assumptions, i.e., to understand under what conditions such a
transceiver structure leads to optimal throughput performance. It
is certainly not obvious that this structure could ever be
optimal, but we show that it is optimal for systems with weak
interference, a situation encountered in many deployed systems
such as DSL.

Our approach in characterizing the sum-rate capacity of a two-user
PGIC leverages recent breakthroughs in determining the sum-rate
capacity of the GIC under noisy interference
\cite{Shang-etal:09IT,Motahari&Khandani:08IT_submission,Annapureddy&Veeravalli:08IT_submission}.
We determine conditions on the channel gains and power constraints
such that there is no loss in terms of sum rate when we impose the
above two assumptions. This is accomplished in two steps. First,
under the independent transmission assumption, we find conditions
such that the maximum sum rate can be achieved by treating
interference as noise in each sub-channel. The key to establishing
these conditions is the concavity of sum-rate capacity in power
constraints for a GIC (cf. Lemma \ref{lemma:PGIC_sumConcave}).
Second, we show that with the same power constraints and channel
gains obtained in the first step, independent transmission and
single-user detection in each sub-channel achieves the sum-rate
capacity of the PGIC. The proof utilizes a genie-aided approach
that generalizes that of \cite{Shang-etal:09IT}.

This paper is organized as follows. In Section \ref{section:model}
we introduce the system model and review recent results. In
Section \ref{section:reformulation}, we consider the maximum sum
rate of a special transmission scheme, i.e., independent
transmission and single-user detection for each sub-channel. We
obtain conditions on the power constraints and channel
coefficients under which the above strategy maximizes the total
sum rate. We prove in Section \ref{section:sumcapacity} that the
maximum sum rate we obtain is the sum-rate capacity. Numerical
results are given in Section \ref{section:numerical}. Section
\ref{section:conclusion} concludes the paper.

\section{System Model and Preliminaries}
\label{section:model}

The received signals of the $i$th  sub-channel $i=1,\cdots,m$ are
defined as
\bqa%
\begin{array}{c}
  Y_{1i}=\sqrt{c_i}X_{1i}+\sqrt{a_i}X_{2i}+Z_{1i}, \\
  Y_{2i}=\sqrt{d_i}X_{2i}+\sqrt{b_i}X_{1i}+Z_{2i}, \\
\end{array}%
\label{eq:PGIC_channel}%
\eqa
where $0\leq a_i<d_i,0\leq b_i<c_i$; $Z_{1i}$ and $Z_{2i}$ are
unit variance Gaussian noise, the total block power constraints
are $P$ and $Q$ for users $1$ and $2$ respectively:
\bqn
\sum_{i=1}^m\left[\frac{1}{n}\sum_{j=1}^nE\left(X_{1i\_j}^2\right)\right]\leq
P,%
\eqn
 and
\bqn%
\sum_{i=1}^m\left[\frac{1}{n}\sum_{j=1}^nE\left(X_{2i\_j}^2\right)\right]\leq
Q,%
\eqn
where $n$ is the block length, and $X_{1i\_j}$ and $X_{2i\_j}$,
$j=1,\dots,n,$ are the user/channel input sequences for the $i$th
sub-channel. We remark that this model is a special case of the
multiple-input multiple-output (MIMO) GIC
\cite{Shang-etal:08Allerton}. We denote the sum-rate capacity of
the $i$th sub-channel as $C_i\left(P_i,Q_i\right)$, where $P_i$
and $Q_i$ are the respective powers allocated to the two users in
this sub-channel.

To find the sum-rate capacity of the PGIC, we need to solve three
problems: the first problem is whether the sub-channels can be
treated separately like the parallel Gaussian multiple-access
channel \cite{Tse&Hanly:98IT} and parallel Gaussian broadcast
channel \cite{Elgamel:80Problemy,Hughes-Hartog:thesis,Tse:98net},
i.e., whether the sum-rate capacity of the PGIC is in the form of
$\sum_{i=1}^mC_i(P_i,Q_i)$. Such a strategy is suboptimal for
PGICs in general
\cite{Cadambe&Jafar:08IT_submission,Sankar-etal:08Allerton}. The
second problem is the optimal distribution of the input signals.
It has been shown respectively in
\cite{Carleial:75IT,Sato:81IT,Han&Kobayashi:81IT} and
\cite{Shang-etal:09IT,Motahari&Khandani:08IT_submission,Annapureddy&Veeravalli:08IT_submission}
that Gaussian inputs are sum-rate optimal for a single-channel GIC
under strong or noisy interference. However, whether this is still
the case for PGICs is not known. The third problem is to find the
optimal power allocation among sub-channels. Existing works on
this problem treat the sub-channels separately, they use Gaussian
inputs, and they use single-user detection at the receivers
\cite{Yu-etal:02JSAC,Song-etal:02COMMAG,Cendrillon-etal:06COM}.

Before proceeding, we introduce the following notation.
\bi%
\item Bold fonts $\xp$ and $\Xbf$ denote vectors and matrices
respectively.
\item $\Ibf$ denotes the identity matrix and $\0bf$ denotes the
zero matrix. \item $|\Xbf|$, $\Xbf^T$, $\Xbf^{-1}$, denote the
respective determinant, transpose and inverse of the matrix
$\Xbf$.
\item $\xp^n=\left[\xp_1^T,\xp_2^T,\dots,\xp_n^T\right]^T$ is a
long vector which consists of the vectors $\xp_i, i=1,\dots, n$.
\item $\xp\sim\Nmat\left(\0bf,\Sigmabf\right)$ means that the
random vector $\xp$ is Gaussian distributed with zero mean and
covariance matrix $\Sigmabf$.
\item $E(\cdot)$ denotes expectation; $\Var(\cdot)$ denotes
variance; $\textrm{Cov}(\cdot)$ denotes covariance matrix;
$I(\cdot;\cdot)$ denotes mutual information; $h(\cdot)$ denotes
differential entropy with the logarithm base $e$ and
$\log(\cdot)=\log_e(\cdot)$.

\ei

\subsection{Noisy-interference sum-rate capacity}

The noisy-interference sum-rate capacity for single-channel GICs
\cite{Shang-etal:09IT,Motahari&Khandani:08IT_submission,Annapureddy&Veeravalli:08IT_submission}
is summarized as follows. \begin{lemma} The sum-rate capacity of
the $i$th sub-channel with $a_i<c_i, b_i<d_i$ and power allocation
$p$, $q$ is
\bqa%
C_i(p,q)=\frac{1}{2}\log\left(1+\dfrac{c_ip}{1+a_iq}\right)+\dfrac{1}{2}\log\left(1+\dfrac{d_iq}{1+b_ip}\right)%
\eqa
provided $(p,q)\in\Amat_i$:
\bqa%
\Amat_i=\left\{\left(\tilde p,\tilde
q\right)\left|\begin{array}{c}
  \sqrt{a_ic_i}(1+b_i\tilde p)+\sqrt{b_id_i}(1+a_i\tilde q)\leq\sqrt{c_id_i} \\
  \tilde p\geq
0,\quad \tilde q\geq 0 \\
\end{array}\right.\right\}.%
\label{eq:PGIC_NI}%
\eqa
   In the case of a
symmetric GIC, i.e., $a_i=b_i, c_i=d_i$ and $p=q$, the noisy
interference condition reduces to \bqa
\frac{a_i}{c_i}\leq\frac{1}{4},\quad
p=q\leq\frac{\sqrt{a_ic_i}-2a_i}{2a_i^2}.\eqa In the case of a ZIC
where $a_i=0$, the noisy interference condition reduces to \bqa
b_i<1,\quad p\geq 0,\quad q\geq 0.\eqa
\label{lemma:PGIC_NIsumCapacity}
\end{lemma}

The main difficulty in maximizing
$\sum_{i=1}^mC_i\left(P_i,Q_i\right)$ is that
$C_i\left(P_i,Q_i\right)$ is generally unknown if
$(P_i,Q_i)\notin\Amat_i$. To solve this problem we use the
following results.

 \vspace{-.2in}
\subsection{Concavity of sum-rate capacity}
 The key to our study of the PGIC is the {\em
concavity} of the sum-rate capacity as a function of the power
constraint. We establish a slightly more general result by using a
modified frequency division multiplexing (FDM) argument
\cite{Sato:78IT}.
%
%
\begin{lemma}%
Let $C_\mu(p,q)$ denote the weighted sum rate capacity of a GIC
with powers $p$ and $q$:
\bqn%
C_\mu(p,q)=\max_{R_1,R_2 \textrm{ achievable}}\{R_1+\mu R_2\},%
\eqn
where $\mu\geq 0$ is a constant. Then $C_\mu(p,q)$ is concave in
the powers $(p,q)$, i.e., for any $0\leq\lambda\leq 1$ we have
\bqa%
 C_\mu(p,q)\geq\lambda
C_\mu(p^\prime,q^\prime)+(1-\lambda)C_\mu(p^{\prime\prime},q^{\prime\prime}),%
\label{eq:PGIC_concave}%
\eqa
where $p^\prime$, $p^{\prime\prime}$, $q^\prime$, and
$q^{\prime\prime}$ are chosen to satisfy
\bqa%
\lambda p^\prime+(1-\lambda)p^{\prime\prime}=p,\quad\lambda
q^\prime+(1-\lambda)q^{\prime\prime}=q. \label{eq:PGIC_TS}%
\eqa
\label{lemma:PGIC_sumConcave}
\end{lemma}
\vspace{-.2in}%
 \bpf
Consider a potentially suboptimal strategy that divides the total
frequency band into two sub-bands: one with a fraction $\lambda$
and the other with a fraction $1-\lambda$ of the total bandwidth.
Powers are allocated into these two sub-bands as $(\lambda
p^\prime,\lambda q^\prime)$ and
$((1-\lambda)p^{\prime\prime},(1-\lambda)q^{\prime\prime})$, where
$ p^\prime,q^\prime,p^{\prime\prime},q^{\prime\prime}$ are such
that (\ref{eq:PGIC_TS}) is satisfied. The information transmitted
in these two sub-bands is independent and the decoding is also
independent. Then the maximum weighted sum rate for the first
sub-band is reduced by a factor $\lambda$ and becomes $\lambda
C_\mu(p^\prime,q^\prime)$. Similarly, the maximum weighted sum
rate at the second sub-band is
$(1-\lambda)C_\mu(p^{\prime\prime},q^{\prime\prime})$. Therefore,
the right-hand side of (\ref{eq:PGIC_concave}) is an achievable
weighted sum rate.%
\epf

Lemma \ref{lemma:PGIC_sumConcave} provides a fundamental result
for weighted sum-rate capacities. It applies not only to two-user
GICs but also to many-user GICs, Gaussian multiaccess channels,
and Gaussian broadcast channels.

\subsection{Subgradient and subdifferential}

To apply the concavity of sum-rate capacity, we need to use
several properties of subgradients and subdifferentials  (see
\cite{Bertsekas:03book}).
\begin{definition}
 If
$f:\Rmat^n\rightarrow \Rmat$ is a real-valued concave function
defined on a convex set $\Smat\subset\Rmat^n$, a vector $\yp$ is a
{\em subgradient} at  point $\xp_0$ if \bqa
f\left(\xp\right)-f\left(\xp_0\right)\leq
\yp^T\left(\xp-\xp_0\right),\quad
\forall\hspace{.1in}\xp\in\Smat.\label{eq:subg}\eqa
\label{def:subg}
\end{definition}
\begin{definition}
For the concave function $f$ defined in Definition \ref{def:subg},
the collection $\partial f\left(\xp_0\right)$ of all subgradients
at point $\xp_0$ is the {\em subdifferential} at this point.
\end{definition}

If the function $f$ is differentiable at $\xp_0$, then the
subgradient and subdifferential both coincide with the gradient.
We introduce a lemma related to subdifferentials which we use to
prove our main result.
\begin{lemma}
Let $f_i(\xp)$, $i=1,\dots, m$, be finite, concave, real-valued
functions on
$\Smat\subset\Rmat^n$ and let $\xp_i^*\in\Smat$, $i=1,\cdots,m$. 
If there is a vector $\yp$ such that $\yp\in\partial
f_i(\xp_i^*),i=1,\dots m$, and $\sum_{i=1}^m\xp_i^*=\up$, then
$\xp^*=\left[{\xp_1^*}^T,\cdots,{\xp_m^*}^T\right]$ is a solution
for the following optimization problem: \bqa
\max&&\quad\sum_{i=1}^mf_i(\xp_i)\nn\\
\textrm{subject
to}&&\quad\sum_{i=1}^m\xp_i=\up,\quad\xp_i\in\Smat.\label{eq:PGIC_optmization}\eqa
\label{lemma:PGIC_subg}
\end{lemma}
\bpf %
Since $\yp\in\partial f_i(\xp_i^*)$, $i=1,\dots m$, we have
\bqa%
 f_i(\xp)\leq
f_i(\xp_i^*)+\yp^T(\xp-\xp_i^*),\quad\forall\hspace{.05in}\xp\in\Smat.
\label{eq:PGIC_subgIn}%
\eqa
%
Let $\hat\xp_i, i=1,\dots,m,$ be any vectors satisfying
$\hat\xp_i\in\Smat$ and $\sum_{i=1}^m\hat\xp_i=\up$, then using
(\ref{eq:PGIC_subgIn}) we have
\bqa%
 f_i(\hat\xp_i)\leq
f_i(\xp_i^*)+\yp^T(\hat\xp_i-\xp_i^*)%
\eqa
 Therefore
 \bqa%
\sum_{i=1}^mf_i(\hat\xp_i)&{}\leq{}&\sum_{i=1}^mf_i(\xp^*_i)+\yp^T\left(\sum_{i=1}^m\hat\xp_i-\sum_{i=1}^m\xp_i^*\right)\nn\\
&{}={}&\sum_{i=1}^mf_i(\xp^*),%
\eqa
 where the last equality is from
$\sum_{i=1}^m\hat\xp_i=\sum_{i=1}^m\xp_i^*=\up$. %
\epf

\begin{figure}[htp]
\centerline{\leavevmode \epsfxsize=5.5in \epsfysize=2.5in
\epsfbox{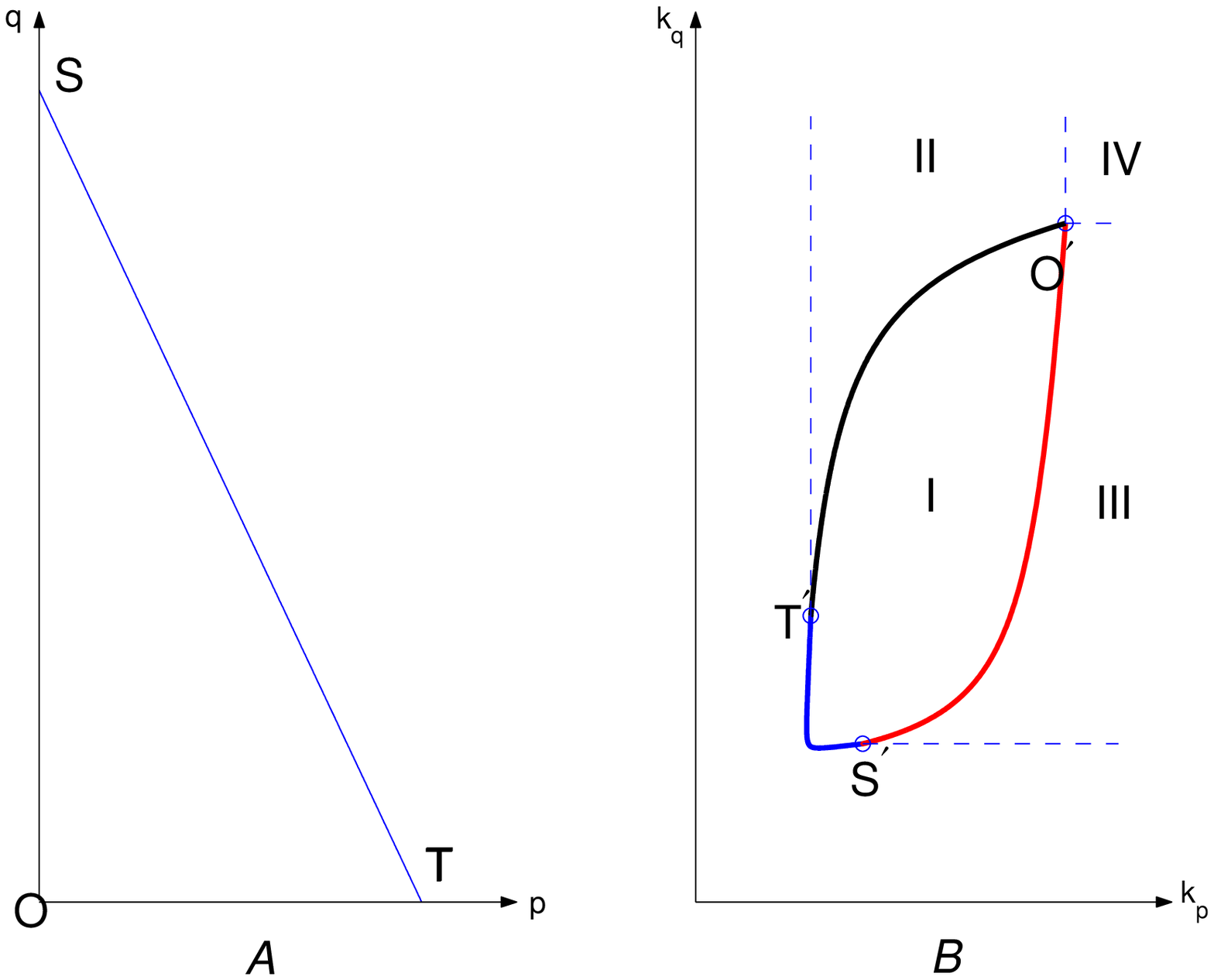}}\caption{The mapping from $\Amat_i$ to
$\Bmat_i$.} \label{fig:mapping}\end{figure}

In the Appendix, we compute the subdifferential $\partial
C_i(p,q)$ when $(p,q)\in\Amat_i$. We are also interested in the
set of pairs
\bqa%
\Bmat_i=\bigcup_{(p,q)\in\Amat_i}\partial C_i(p,q).%
\eqa
The mapping from $\Amat_i$ to $\Bmat_i$ is illustrated in Fig.
\ref{fig:mapping}. As seen from (\ref{eq:PGIC_NI}), $\Amat_i$ is a
triangle region with the corner points
\bqn%
 &&O(0,0),\\
 &&S\left(0,\dfrac{\sqrt{c_id_i}-\sqrt{a_ic_i}-\sqrt{b_id_i}}{a_i\sqrt{b_id_i}}\right)\triangleq(0,q_s),%
\eqn
and
\bqn%
 &&T\left(\dfrac{\sqrt{c_id_i}-\sqrt{a_ic_i}-\sqrt{b_id_i}}{b_i\sqrt{a_ic_i}},0\right)\triangleq(p_t,0).
\eqn
 The corresponding points in $\Bmat_i$ are respectively
\bqn%
&& O^\prime\left(\dfrac{c_i}{2},\dfrac{d_i}{2}\right),\\
&&
S^\prime\left(\dfrac{c_i}{2(1+a_iq_s)}-\dfrac{b_id_iq_s}{2(1+d_iq_s)},\dfrac{d_i}{2(1+d_iq_s)}\right),\\
\eqn
and
\bqn%
&&T^\prime\left(\dfrac{c_i}{2(1+c_ip_t)},\dfrac{d_i}{2(1+b_ip_t)}-\dfrac{a_ic_ip_t}{2(1+c_ip_t)}\right).
\eqn
%

 Let $\Amat_i^{(1)}$ be
the inner points of $\Amat_i$ and the line segment
$\overline{ST}$, and let $\Bmat_i^{(1)}$ be the inner points of
the closed area defined by $O^\prime S^\prime T^\prime$ and the
curve $\widehat{S'T'}$. As shown in the Appendix, $\Amat_i^{(1)}$
maps to $\Bmat_i^{(1)}$ and this mapping is one-to-one. Let
$\Amat_i^{(2)}$ be the line segment $\overline{OT}$ and
$\Bmat_i^{(2)}$ be the curve $\widehat{O'T'}$ and the points above
it (labeled as region II in Fig. \ref{fig:mapping}).
$\Amat_i^{(2)}$ maps to $\Bmat_i^{(2)}$ and this mapping is
one-to-many. Specifically, let $(P_i,0)$ be a point on
$\overline{OT}$. The partial derivatives of $C_i(p,q)$ with
respect to $p$ and $q$ at this point are denoted as $K_p$ and
$K_q$, respectively, where $K_p$ is a two-sided partial derivative
and $K_q$ is a one-sided partial derivative. Point $(K_p,K_q)$ is
on the curve $\widehat{O^\prime T^\prime}$ of $\Bmat_i^{(2)}$. The
subdifferential of $C_i(p,q)$ at point $(P_i,0)$ is a ray in
$\Bmat_i^{(2)}$ defined as $k_p=K_p$, $k_q\geq K_q$.

Similarly to the above, let $\Amat_i^{(3)}$ be the line segment
$\overline{OS}$ and $\Bmat_i^{(3)}$ be the curve $\widehat{O'S'}$
and the points to the right (labeled as region III in Fig.
\ref{fig:mapping}). $\Amat_i^{(3)}$ maps to $\Bmat_i^{(3)}$ and
this mapping is also one-to-many. Let $\Amat_i^{(4)}$ be the
origin and let $\Bmat_i^{(4)}$ be the collection of points
$(k_p,k_q)$ satisfying  $k_p\geq K_p$ and $k_q\geq K_q$ (labeled
as region IV in Fig. \ref{fig:mapping}), where $K_p$ and $K_q$ are
the two one-sided partial derivatives at the origin.
$\Amat_i^{(4)}$ maps to $\Bmat_i^{(4)}$.


\subsection{Concave-like property of conditional entropy}
The following Lemma is proved in \cite{Shang-etal:08Allerton}
based on the fact that a Gaussian distribution maximizes
conditional entropy under a covariance matrix constraint
\cite{Thomas:87IT}.
\begin{lemma}\cite[Lemma 2]{Shang-etal:08Allerton}%
${}$ Let
$\xp_i^n=\left[\xp_{i,1}^T,\dots,\xp_{i,n}^T\right]^T,i=1,\dots,k$,
be $k$ long random vectors each of which consists of $n$ vectors.
Suppose the $\xp_{i,j}$, $i=1,\cdots,k$ all have the same length
$L_j$, $j=1,\cdots,n$. Let
$\yp^{n}=\left[\yp_1^T,\dots,\yp_n^T\right]^T$, where $\yp_j$ has
length $L_j$, be a long
Gaussian random vector with covariance matrix%
\bqa
\Cov\left(\yp^{n}\right)=\sum_{i=1}^k\lambda_i\Cov\left(\xp^n_i\right),\label{eq:cvxCov}\eqa
where $\sum_{i=1}^k\lambda_i=1,\lambda_i\geq 0$. Let $\Smat$ be a
subset of $\{1,2,\dots,n\}$ and $\Tmat$ be a subset of $\Smat$'s
complement. Then we have
\bqa%
 \sum_{i=1}^k\lambda_i
h\left(\xp_{i,\Smat}\left|\xp_{i,\Tmat}\right.\right)\leq
h\left(\yp_\Smat\left|\yp_{\Tmat}\right.\right).%
\eqa
\label{lemma:generalconcave}
\end{lemma}
When $\xp_k$, $k=1,\cdots,n$ are all Gaussian distributed, Lemma
\ref{lemma:generalconcave} shows that
$h\left(\xp_\Smat\left|\xp_{\bar\Smat}\right.\right)$ is concave
over the covariance matrices.

\section{A lower bound for the sum-rate capacity}
\label{section:reformulation}

If the sum-rate capacity of a PGIC can be achieved by (1)
transmitting independent symbol streams in each sub-channel and
(2) treating interference as noise in each sub-channel, we say
this PGIC has noisy interference. Before proceeding to the main
theorem of noisy-interference sum-rate capacity, we first consider
the following optimization problem:
\bqa%
\max&&\quad\sum_{i=1}^mC_i(P_i,Q_i)\nn\\
\textrm{subject to}&&\quad \sum_{i=1}^mP_i=P, \quad
\sum_{i=1}^mQ_i=Q\nn\\
&&\quad P_i\geq 0,\quad Q_i\geq 0,\quad
i=1,\dots,m.%
\label{eq:PGIC_individualSum}%
\eqa
 Problem
(\ref{eq:PGIC_individualSum}) is to find the maximum of the sum of
the sum-rate capacities of individual sub-channels and the
corresponding power allocation. In general, the optimal solution
of (\ref{eq:PGIC_individualSum}) is {\em not} the sum-rate
capacity of the PGIC, since it presumes that the signals
transmitted in each sub-channel are independent and no joint
decoding across sub-channels is allowed. However, solving problem
(\ref{eq:PGIC_individualSum}) is important to derive the sum-rate
capacity of a PGIC. We are interested in the case where the
optimal power allocations $P_i^*$, $Q_i^*$ satisfy the following
noisy interference conditions \bqa
\sqrt{a_ic_i}(1+b_iP_i^*)+\sqrt{b_id_i}(1+a_iQ_i^*)\leq\sqrt{c_id_i},\quad
i=1,\dots,m.\label{eq:PGIC_individualNI}\eqa In such a case, it
turns out that the sum of the sum-rate capacities in
(\ref{eq:PGIC_individualSum}) is maximized when each sub-channel
experiences noisy interference.

For the rest of this section, we first consider the general PGIC
and derive the optimal solution of problem
(\ref{eq:PGIC_individualSum}) based on Lemmas
\ref{lemma:PGIC_sumConcave} and \ref{lemma:PGIC_subg}. We further
find conditions on the total power $P$ and $Q$ such that the
optimal solution of (\ref{eq:PGIC_individualSum}) satisfies
(\ref{eq:PGIC_individualNI}). Then we focus on symmetric PGICs and
provide some insights on this solution.

\subsection{General parallel Gaussian interference channel}
\begin{theorem}
For a PGIC defined in (\ref{eq:PGIC_channel}),  if
$\sqrt{a_ic_i}+\sqrt{b_id_i}<\sqrt{c_id_i}$ and the power
constraint $(P,Q)$ is in the following set
\bqa
\bigcup_{[k_p^*,k_q^*]^T\in\bigcap_{i=1}^m\Bmat_i}\left\{(P,Q)\left|\begin{array}{l}
  P=\sum_{i=1}^mP_i^*,\quad Q=\sum_{i=1}^mQ_i^*, \\
  {[k_p^*,k_q^*]^T}\in\partial C_i(P_i^*,Q_i^*)\quad i=1,\dots,m. \\
\end{array}\right.\right\}, %
\label{eq:PGIC_generalPower}%
\eqa
then the optimal solution of (\ref{eq:PGIC_individualSum})
satisfies (\ref{eq:PGIC_individualNI}).
\label{thm:pgic}
\end{theorem}

\bpf%
 The proof is straightforward from Lemma
\ref{lemma:PGIC_subg}. For any
$[k_p^*,k_q^*]^T\in\bigcap_{i=1}^m\Bmat_i$, there exist
$P_i^*,Q_i^*$ such that ${[k_p^*,k_q^*]^T}\in\partial
C_i(P_i^*,Q_i^*), i=1,\dots,m$. Thus if
$P=\sum_{i=1}^mP_i^*,Q=\sum_{i=1}^mQ_i^*$, then from Lemma
\ref{lemma:PGIC_subg}, $P_i^*,Q_i^*$ are optimal for the
optimization problem (\ref{eq:PGIC_individualSum}). Since
$(P_i^*,Q_i^*)\in\Amat_i$, then from Lemma
\ref{lemma:PGIC_NIsumCapacity}, $(P_i^*,Q_i^*)$ satisfies
(\ref{eq:PGIC_individualNI}). \epf
%

Theorem \ref{thm:pgic} provides conditions on the power and
channel coefficients such that treating interference as noise (or
single-user detection) maximizes the sum rate of a PGIC under the
assumption of independent transmission among sub-channels. The
conditions of Theorem \ref{thm:pgic} ensures that the power
constraints $P$ and $Q$ are associated with a subgradient
$\left[k_p^*,k_q^*\right]^T$ shared by $C_i(P_i^*,Q_i^*)$ for all
$i=1,\cdots,m$. Therefore, at the points of the optimal power
allocations $(P_i^*,Q_i^*)$, all the functions $C_{i}(p_i,q_i)$
have parallel supporting hyperplanes. We will discuss this in more
details in Remark 1 below.

 In general, the closed-form expression (\ref{eq:PGIC_generalPower}) of the power region
for $P$ and $Q$ is very complex. However, for some special cases
like symmetric PGICs, we can obtain simpler closed-form solutions.

\subsection{Symmetric parallel Gaussian interference channel}%
In this section, we consider PGICs with symmetric parameters,
namely $a_i=b_i,c_i=d_i$ and $P=Q$. Without loss of generality we
assume $c_1\geq c_2\cdots\geq c_m$. Define \bqa
w_i&{}={}&\frac{4a_i^2}{\left(\sqrt{c_i}-\sqrt{a_i}\right)^2},\\
\hat w&{}={}&\max_{i}\{w_i\},\eqa and let $r$ to be an index
between $1$ and $m$ such that
\bqa %
c_{r+1}<\hat w< c_r,%
\label{eq:r}%
\eqa
where we let $c_{m+1}=0$ for convention. Then
we have the following theorem.
\begin{theorem} For a symmetric PGIC, if $\dfrac{a_i}{c_i}<\dfrac{1}{4}$, $i=1,\dots,m$, and
\bqa%
0<P\leq\bar P,%
\label{eq:symmPrange}%
\eqa
where
\bqa%
 \bar
P=\sum_{i=1}^r\frac{\sqrt{c_i^2+\dfrac{4a_ic_i}{\hat
w}(a_i+c_i)}-(2a_i+c_i)}{2a_i(a_i+c_i)},%
\label{eq:barP}%
\eqa
then the optimal solution of (\ref{eq:PGIC_individualSum})
satisfies (\ref{eq:PGIC_individualNI}). Furthermore, only the
first $r$ sub-channels are active. \label{thm:PGIC_symmpgic}
\end{theorem}

\bpf  By symmetry, we simplify the proof by considering the
following optimization problem:
\bqa
\begin{array}{rl}
  \max &\quad \displaystyle\sum_{i=1}^mC_i(P_i,P_i) \\
  \textrm{subject to} &\quad \displaystyle\sum_{i=1}^mP_i=P \\
   &\quad P_i\geq 0, \quad i=1,\dots,m. \\
\end{array}%
\label{eq:PGIC_mainSymm}%
\eqa
That is, we require that the power allocated to both users be
$P_i$ for the $i$th sub-channel. Obviously the maximum of
(\ref{eq:PGIC_mainSymm}) is no greater than the maximum of
$(\ref{eq:PGIC_individualSum})$ because of the extra constraint
$P_i=Q_i$. To prove Theorem \ref{thm:PGIC_symmpgic}, it suffices
to show that under the condition $0<P\leq\bar P$: $1)$ the optimal
$P_i^*$ for (\ref{eq:PGIC_mainSymm}) satisfy the noisy
interference condition; $2)$ the optimization problems
(\ref{eq:PGIC_individualSum}) and (\ref{eq:PGIC_mainSymm}) are
equivalent.

Let $P_i=Q_i$, we obtain from Lemma \ref{lemma:PGIC_NIsumCapacity}
\bqa%
 \Amat_i^\prime=\left\{p\left|0\leq
p\leq\frac{\sqrt{a_ic_i}-2a_i}{2a_i^2}\right.\right\}.%
\label{eq:symmAi}%
\eqa
The subdifferential is computed in the Appendix and is given by
(see (\ref{eq:subd}))
\bqa%
 \partial C_i(P_i)=\left\{\begin{array}{ll}
  \left\{k\left|c_i\leq k\leq \hat c\right.\right\}, & \quad P_i=0, \\
  \left\{k\left|k=\dfrac{c_i}{(1+a_iP_i)(1+a_iP_i+c_iP_i)}\right.\right\}, & \quad P_i\in\Amat_i^\prime,\quad P_i\neq 0, \\
\end{array}\right.%
\label{eq:symmSD}%
\eqa
where $\hat c=\max\{c_i\}$. Therefore
\bqa%
\Bmat_i^\prime=\bigcup_{P_i\in\Amat_i^\prime}\partial
C_i(P_i)=\left\{k\left|w_i\leq k\leq\hat c\right.\right\},%
\eqa
and
\bqa%
 \bigcap_{i=1}^m\Bmat_i^\prime=\left\{k\left|\hat w\leq
k\leq\hat c\right.\right\}.%
\eqa
For any $k^*\in\bigcap_{i=1}^m\Bmat_i^\prime$, equation
(\ref{eq:symmSD}) determines a one-to-one mapping from $k^*$ to
$P_i^*\in\Amat_i^\prime$, namely
\bqa%
P_i^*(k^*)=\left\{\begin{array}{ll}
  0, &\quad k^*\geq c_i \\
  \dfrac{\sqrt{c_i^2+\dfrac{4a_ic_i}{k^*}(a_i+c_i)}-(2a_i+c_i)}{2a_i(a_i+c_i)}, &\quad w_i\leq k^*<c_i. \\
\end{array}\right.\label{eq:symmPistar}%
\eqa
So consider the region (\ref{eq:PGIC_generalPower}) which is here
\bqa%
 \bigcup_{k^*\in\left[\hat w, \hat
c\right]}\left\{P\left|P=\sum_{i=1}^mP_i^*(k^*)\right.\right\}.
\label{eq:symmP}%
\eqa
 From (\ref{eq:symmPistar}), $\sum_{i=1}^mP_i^*(k^*)$ is decreasing in
$k^*$, therefore
\bqa%
 P&{}\geq{}&P(k^*=\hat c)=0,\\
P&{}\leq{}&P(k^*=\hat w)=\sum_{i=1}^rP_i^*(k^*=\hat
w)\nn\\
&=&\sum_{i=1}^r\frac{\sqrt{c_i^2+\dfrac{4a_ic_i}{\hat
w}(a_i+c_i)}-(2a_i+c_i)}{2a_i(a_i+c_i)}\triangleq\bar
P,%
\label{eq:maxSymmP}%
\eqa
 where the first equality of
(\ref{eq:maxSymmP}) is from (\ref{eq:r}). Since
$P(k^*)=\sum_{i=1}^mP_i^*(k^*)$ is continuous over $k^*$, for any
$P\in[0,\bar P]$ there exists a $k^*$, and the corresponding
$P_i^*, i=1,\dots,m,$ that solve the optimization problem
(\ref{eq:PGIC_mainSymm}).

We complete the proof by showing that the optimal $P_i^*$ for
(\ref{eq:PGIC_mainSymm}) is also optimal for
(\ref{eq:PGIC_individualSum}) for a symmetric PGIC. Assume that
for a given $P$ and the optimal $P_i^*$ of
(\ref{eq:PGIC_mainSymm}), the corresponding subgradient (which is
identical for all $i$) is $k^*$. Then by symmetry, the
subderivative of $C_i(P_i,Q_i)$ in (\ref{eq:PGIC_individualSum})
is $[\frac{k^*}{2},\frac{k^*}{2}]^T$ by choosing $P_i=Q_i=P_i^*$.
Therefore the subderivatives are identical for all the
$C_i(P_i,Q_i)$ at $P_i=Q_i=P_i^*$. From Lemma
\ref{lemma:PGIC_subg}, $P_i=Q_i=P_i^*$ is an optimal choice for
(\ref{eq:PGIC_individualSum}). Since $P_i^*$ satisfies the
noisy-interference condition in (\ref{eq:symmAi}), $P_i=Q_i=P_i^*$
also satisfies the noisy-interference condition in
(\ref{eq:PGIC_NI}). \epf

\begin{figure}[htp]
\centerline{\leavevmode \epsfxsize=5in \epsfysize=4in
\epsfbox{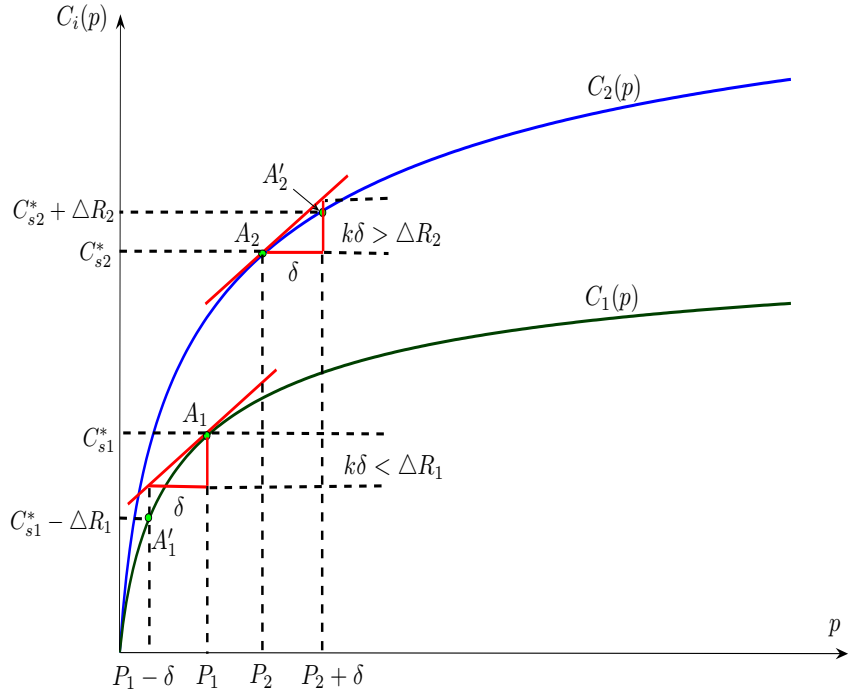}}\caption{An illustration of sum-rate
capacity achieving power allocation for a symmetric parallel
Gaussian interference channel.}
\label{fig:PGIC_illustration}\end{figure}

 {\em Remark 1}: From the proof of Theorem \ref{thm:PGIC_symmpgic}, all the $C_i(P_i,Q_i)$ have parallel supporting
 hyperplanes at the optimal point $P_i=Q_i=P_i^*$. This gives rise to a geometric interpretation, as illustrated
 in Fig. \ref{fig:PGIC_illustration}. For clarity, we use the simplified optimization problem (\ref{eq:PGIC_mainSymm}).
 $C_i(p)$ is the sum-rate capacity for the $i$th sub-channel, points $A_1$ and
 $A_2$ correspond to the power allocations $P_1$ and $P_2$, respectively, and the two supporting hyperplanes pass through $A_1$ and $A_2$. The power allocation
 satisfies $P=P_1+P_2, P_i\in\Amat_i^\prime, i=1,2$, and $k=\frac{\partial C_1(p)}{\partial p}\left|_{p=P_1}\right.=\frac{\partial C_2(p)}{\partial
 p}\left|_{p=P_2}\right.$ (we assume the subgradient is equal to the gradient in this case, and hence the supporting hyperplane is the tangent hyperplane),
 and the corresponding sum rate is $C_{s1}^*+C_{s2}^*$. Consider
 the
 power allocation $P_1-\delta$ and $P_2+\delta$
 and the corresponding sum-rate capacities for the two sub-channels
 $C_{s1}^*-\triangle{R_1}$ and $C_{s2}^*+\triangle{R_2}$, respectively.
 By
 concavity, $\triangle{R_1}>k\delta$ and $\triangle{R_2}<k\delta$.
 Therefore the new sum-rate is $\left(C_{s1}^*-\triangle{R_1}\right)+\left(C_{s2}^*+\triangle{R_2}\right)
 <\left(C_{s1}^*-k\delta\right)+\left(C_{s2}^*+k\delta\right)=C_{s1}^*+C_{s2}^*$.

{\em Remark 2}: When
${\displaystyle\min_{i}}\left\{\frac{\sqrt{a_ic_i}-2a_i}{2a_i^2}\right\}<P\leq\bar
P$, there exist power allocations such that some sub-channels do
not have noisy interference. As such the sum-rate capacities of
those sub-channels are unknown. Surprisingly, in this case we do
not need to derive upper bounds for those unknown sum-rate
capacities. Instead, the concavity of the sum-rate capacity (as a
function of the power) and the existing noisy-interference
sum-rate capacity results ensure the validity of Theorem
\ref{thm:PGIC_symmpgic}.

{\em Remark 3}: The parallel supporting hyperplanes condition for
the optimal power allocation is applicable to a broad class of
parallel channels in which 1) transmissions across subchannels are
independent, 2) the capacity of each subchannel is concave in its
power constraint. For example, this condition applies to parallel
multi-access and broadcast channels. In particular, applying the
condition to single user parallel Gaussian channels, it is easy to
verify that the parallel supporting hyperplanes condition reduces
to the classic waterfilling interpretation.

{\em Remark 4}: Intuitively, since each sub-channel is a symmetric
Gaussian IC with noisy interference, the power allocated to the
two users in each sub-channel ought to be identical. While Theorem
\ref{thm:PGIC_symmpgic} does not explicitly address the power
allocation scheme, we see from the proof of the theorem that this
is indeed the case.

{\em Remark 5}: If (\ref{eq:symmPrange}) is satisfied, the optimal
power allocation $P_i^*$ is unique, and there exists a
$k^*\in[\hat w,\hat c]$ such that $P_i^*$ and $k^*$ satisfy
(\ref{eq:symmPistar}). To see this, observe that in the proof of
Theorem \ref{thm:PGIC_symmpgic},
$\sum_{i=1}^mP_i^*\left(k^*\right)$ is continuous and
monotonically decreasing in $k^*$ when $k^*\in[\hat w,\hat c]$,
and $P$ varies from $0$ to $\bar P$ when $k^*$ varies from $\hat
c$ to $\hat w$. Thus, if $0\leq P\leq\bar P$, there exists a
corresponding unique $k^*$ in $[\hat w, \hat c]$ that solves
problem (\ref{eq:PGIC_mainSymm}). Since the mapping from $k^*$ to
$P_i^*$ in (\ref{eq:symmPistar}) is a one-to-one mapping, $P_i^*$
is also unique.


{\em Remark 6}: As shown in (\ref{eq:symmPistar}), whether a
sub-channel is active or not depends only on the direct channel
gain $c_i$. The amount of power allocated to a sub-channel depends
on both the direct channel gain $c_i$ and the interference channel
gain $a_i$. When the total power constraint $P$ increases from $0$
to $\bar P$, the corresponding $k^*$ decreases from $\hat c$ to
$\hat w$. As such, from (\ref{eq:symmPistar}) the sub-channels
with larger $c_i$ become active earlier than those with smaller
$c_i$.

\section{Noisy interference sum-rate capacity}
\label{section:sumcapacity}

The following theorem gives the noisy-interference sum-rate
capacity of a PGIC.

\begin{theorem}
 For the PGIC defined in (\ref{eq:PGIC_channel}), if
\bqa\sqrt{a_ic_i}+\sqrt{b_id_i}<\sqrt{c_id_i}\label{eq:PGIC_chcoefficient}\eqa
for all $i=1,\cdots,m$, and the power constraint pair $(P,Q)$ is
in the set (\ref{eq:PGIC_generalPower}), the sum-rate capacity is
the maximum of problem (\ref{eq:PGIC_individualSum}), and the
sum-rate capacity is achieved by independent transmission across
sub-channels and treating interference as noise for each
sub-channel.
\label{theorem:PGIC}
\end{theorem}

The following theorem is a special case of Theorem \ref{thm:pgic}
for symmetric PGICs.
\begin{theorem}
For a symmetric PGIC, if $\dfrac{a_i}{c_i}<\dfrac{1}{4}$ for all
$i=1,\cdots,m$, and the power constraint $P$ satisfies
(\ref{eq:symmPrange}), then the sum-rate capacity is the maximum
of problem (\ref{eq:PGIC_mainSymm}) and is
 achieved by independent transmission across sub-channels and treating interference as noise in
each sub-channel.\label{theorem:symmPGIC}
\end{theorem}

For the PGIC, there may exist some sub-channels with one-sided
interference or no interference, i.e., $b_j=0$, or $a_k=0$ or
$a_r=b_r=0$ for some integers $j,k$ and $r$ between $1$ and $m$.
We prove Theorem \ref{theorem:PGIC} for all such cases.

\bpf Let $i,j,k,r$ be integers and $1\leq i\leq m_1$, $m_1+1\leq
j\leq m_2$, $m_2+1\leq k\leq m_3$ and $m_3+1\leq r\leq m$
throughout this proof. We denote $\mi,\mj,\mk,\mr$ as index sets
and $\mi=\{1,\cdots,m_1\}$, $\mj=\{m_1+1,\cdots,m_2\}$,
$\mk=\{m_2+1,\cdots,m_3\}$, and $\mr=\{m_3+1,\cdots,m\}$. Without
loss of generality we can assume that sub-channels with index $i$
are all two-sided GICs with $a_i\neq 0$, $b_i\neq 0$ ;  the
sub-channels with index $j$ are all GICs with $a_j\neq 0$,
$b_j=0$; the sub-channels with index $k$ are all GICs with
$a_k=0$, $b_k\neq 0$; and the sub-channels with index $r$ are all
GICs with $a_r=b_r=0$. Let
$\Abf=\diag(\sqrt{a_1},\cdots,\sqrt{a_m})$,
$\Bbf=\diag(\sqrt{b_1},\cdots,\sqrt{b_m})$,
$\Cbf=\diag(\sqrt{c_1},\cdots,\sqrt{c_m})$,
$\Dbf=\diag(\sqrt{d_1},\cdots,\sqrt{d_m})$. Then we can rewrite
(\ref{eq:PGIC_channel}) in the following form %
\bqa%
\begin{array}{c}
  \yp_1=\Cbf\xp_1+\Abf\xp_2+\zp_1, \\
  \yp_2=\Bbf\xp_1+\Dbf\xp_2+\zp_2, \\
\end{array}%
\label{eq:MIMO}%
\eqa%
where $\zp_1\sim\Nmat\left(\0bf,\Ibf\right)$ and
$\zp_2\sim\Nmat\left(\0bf,\Ibf\right)$.

We further define%
 \bqn%
\Abf_{\underline{i}}&{}={}&\diag(\sqrt{a_1},\cdots,\sqrt{a_{m_1}})=\diag(\sqrt{a_\mi}),\\
\Abf_{\underline{j}}&{}={}&\diag(\sqrt{a_{m_1+1}},\cdots,\sqrt{a_{m_2}})=\diag(\sqrt{a_\mj}),\\
\Abf_{\underline{k}}&{}={}&\diag(\sqrt{a_{m_2+1}},\cdots,\sqrt{a_{m_3}})=\diag(\sqrt{a_\mk}),\\
\Abf_{\underline{r}}&{}={}&\diag(\sqrt{a_{m_3+1}},\cdots,\sqrt{a_{m}})=\diag(\sqrt{a_\mr}),\eqn
and similarly for $\Bbf$, $\Cbf$ and $\Dbf$. Denote the
transmitted vector of user 1 as $\xp_1=[X_1,\cdots,X_m]$, where
each entry is the transmitted signal at the corresponding
sub-channel. Similarly, we let
$\xp_{1\underline{i}}=[X_1,\cdots,X_{m_1}]=[X_{1\mi}]$,
$\xp_{1\underline{j}}=[X_{m_1+1},\cdots,X_{m_2}]=[X_{1\mj}]$,
$\xp_{1\underline{k}}=[X_{m_2+1},\cdots,X_{m_3}]=[X_{1\mk}]$ and
$\xp_{1\underline{r}}=[X_{m_3+1},\cdots,X_{m}]=[X_{1\mr}]$. The
input vectors for the second user are similarly defined.

Since the power constraint $[P,Q]^T$ is in the set
(\ref{eq:PGIC_generalPower}), so there exists a subgradient
$[k_p^*,k_q^*]^T\in\bigcap_{l=1}^m\Bmat_l$ and the corresponding
$\left[P_l^*,Q_l^*\right]^T\in\Amat_l$ such that
\bqa%
[k_p^*,k_q^*]^T\in\partial C_l\left(P_l^*,Q_l^*\right),\quad
l=1,\cdots,m. \label{eq:identicalsg}%
\eqa
From Theorem \ref{thm:pgic}, the $P_l^*,Q_l^*$ optimize problem
(\ref{eq:PGIC_individualSum}).

Assuming the channel is used $n$ times, the transmitted vector
sequences are denoted as
$\xp_1^n=\left[\xp_{11}^T,\cdots,\xp_{1n}^T\right]^T$ and
$\xp_2^n=\left[\xp_{21}^T,\cdots,\xp_{2n}^T\right]^T$ which
satisfy the average power constraints
\bqa%
&&\sum_{l=1}^n\tr\left[E\left(\xp_{1l}\xp_{1l}^T\right)\right]\leq
nP,\nn\\%
&&\sum_{l=1}^n\tr\left[E\left(\xp_{2l}\xp_{2l}^T\right)\right]\leq
nQ.\nn
\eqa%
Define zero-mean Gaussian vectors $\widehat\xp_1^*=\left[\widehat
X_{11}^*,\cdots,\widehat X_{1m}^*\right]$ and
$\widehat\xp_2^*=\left[\widehat X_{21}^*,\cdots,\widehat
X_{2m}^*\right]$ with the covariance matrices %
\bqa%
&&\Cov\left(\widehat\xp_1^*\right)=\frac{1}{n}\sum_{l=1}^n\Cov\left(\xp_{1l}\right),\label{eq:x1hat}\\
&&\Cov\left(\widehat\xp_2^*\right)=\frac{1}{n}\sum_{l=1}^n\Cov\left(\xp_{2l}\right).\label{eq:x2hat}
\eqa%
Obviously, $\widehat\xp_1^*$ and $\widehat\xp_2^*$ satisfy the
power constraints. We define
\bqa%
P_l&{}={}&\Var\left(\widehat X_{1l}^*\right),\\
Q_l&{}={}&\Var\left(\widehat X_{2l}^*\right).%
\eqa
Vectors $\widehat\yp_1^*$ and $\widehat\yp_2^*$ are defined by
(\ref{eq:MIMO}) with $\xp_1$ and $\xp_2$ being replaced by
$\widehat\xp_1^*$ and $\widehat\xp_2^*$ respectively. Similar to
$\xp_1$, $\widehat\xp_{1}^*$ is also partitioned as
$\widehat\xp_{1\mi}^*$, $\widehat\xp_{1\mj}^*$,
$\widehat\xp_{1\mk}^*$ and $\widehat\xp_{1\mr}^*$.

Define Gaussian random vectors $\np_{1\mi},\np_{1\mj},\np_{2\mi}$
and $\np_{2\mk}$ independent of $\xp_1$ and $\xp_2$, and let
\bqa
&&\left[\begin{array}{c}
  \zp_{1\mi} \\
  \zp_{1\mj} \\
  \zp_{1\mk} \\
  \np_{1\mi} \\
  \np_{1\mj} \\
\end{array}\right]\sim\Nmat\left(\0bf,\left[\begin{array}{ccccc}
  \Ibf_\mi & \0bf &\quad \0bf &\quad \rho_{1\mi}\sigma_{1\mi} &\quad \0bf \\
  \0bf & \Ibf_\mj &\quad \0bf &\quad \0bf &\quad \rho_{1\mj}\sigma_{1\mj} \\
  \0bf & \0bf &\quad \Ibf_\mk &\quad \0bf &\quad \0bf \\
  \rho_{1\mi}\sigma_{1\mi} & \0bf & \quad\0bf &\quad \sigma_{1\mi}^2 & \quad\0bf \\
  \0bf & \rho_{1\mj}\sigma_{1\mj} & \quad\0bf &\quad \0bf &\quad \sigma_{1\mj}^2 \\
\end{array}\right]\right),\label{eq:genie1}
\\
&&\left[\begin{array}{c}
  \zp_{2\mi} \\
  \zp_{2\mj} \\
  \zp_{2\mk} \\
  \np_{2\mi} \\
  \np_{2\mk} \\
\end{array}\right]\sim\Nmat\left(\0bf,\left[\begin{array}{ccccc}
  \Ibf_\mi & \0bf &\quad \0bf &\quad \rho_{2\mi}\sigma_{2\mi} &\quad \0bf \\
  \0bf & \Ibf_\mj &\quad \0bf &\quad \0bf &\quad  \0bf\\
  \0bf & \0bf &\quad \Ibf_\mk &\quad \0bf &\quad \rho_{2\mk}\sigma_{2\mk} \\
  \rho_{2\mi}\sigma_{2\mi} & \0bf & \quad\0bf &\quad \sigma_{2\mi}^2 & \quad\0bf \\
  \0bf & \0bf & \quad\rho_{2\mk}\sigma_{2\mk} &\quad \0bf &\quad \sigma_{2\mk}^2 \\
\end{array}\right]\right),\label{eq:genie2}
\eqa
where $\rho_{\mi}$ and $\sigma_{\mi}$ are diagonal matrices with
 the diagonal entries being $\rho_i$ and $\sigma_i$, $i\in\mi$, respectively. Furthermore, we let
\bqa
\sigma_{1i}^2&{}={}&\frac{1}{2b_i}\left(\frac{b_i}{c_i}\left(a_iQ_i^*+1\right)^2-\frac{a_i}{d_i}\left(b_iP_i^*+1\right)^2+1\right.\nn\\
&&\hspace{.4in}\left.\pm\sqrt{\left[\frac{b_i}{c_i}\left(a_iQ_i^*+1\right)^2-\frac{a_i}{d_i}\left(b_iP_i^*+1\right)^2+1\right]^2-\frac{4b_i}{c_i}\left(a_iQ_i^*+1\right)^2}\right),\label{eq:s1i}\\
\sigma_{2i}^2&{}={}&\frac{1}{2a_i}\left(\frac{a_i}{d_i}\left(b_iP_i^*+1\right)^2-\frac{b_i}{c_i}\left(a_iQ_i^*+1\right)^2+1\right.\nn\\
&&\hspace{.4in}\left.\pm\sqrt{\left[\frac{a_i}{d_i}\left(b_iP_i^*+1\right)^2-\frac{b_i}{c_i}\left(a_iQ_i^*+1\right)^2+1\right]^2-\frac{4a_i}{d_i}\left(b_iP_i^*+1\right)^2}\right),\label{eq:s2i}\\
\rho_{1i}&{}={}&\sqrt{1-a_i\sigma_{2i}^2},\label{eq:rho1i}\\
\rho_{2i}&{}={}&\sqrt{1-b_i\sigma_{1i}^2},\label{eq:rho2i}\\
\sigma_{1j}^2&{}={}&\frac{\left(1+a_jQ_j^*\right)^2}{c_j\rho_{1j}^2},\label{eq:s1j}\\
\rho_{1j}&{}={}&\sqrt{1-\frac{a_j}{d_j}},\label{eq:rho1j}\\
\sigma_{2k}^2&{}={}&\frac{\left(1+b_kP_k^*\right)^2}{d_k\rho_{2k}^2},\label{eq:s2k}\\
\rho_{2k}&{}={}&\sqrt{1-\frac{b_k}{c_k}}.\label{eq:rho2k}
\label{eq:rho2k}%
\eqa
We emphasize that the $P_l^*$ and $Q_l^*$ in
(\ref{eq:s1i})-(\ref{eq:rho2k}) are the {\em optimal} powers for
the problem (\ref{eq:PGIC_individualSum}) and can be considered as
constants in what follows.
 It
has been shown in \cite[equations (51),(52)]{Shang-etal:09IT} that
(\ref{eq:s1i})-(\ref{eq:rho2i}) are feasible (i.e., there exist at
least one choice of
$\{\sigma_{1i}^2,\sigma_{2i}^2,\rho_{1i},\rho_{2i}\}$ such that
the covariance matrices are symmetric and semi-positive definite,
and thus the defined Gaussian random vectors exist) for the
definition in (\ref{eq:genie1}) and (\ref{eq:genie2}) if and only
if $\left[P_i^*,Q_i^*\right]^T\in\Amat_i$. Obviously,
(\ref{eq:s1j})-(\ref{eq:rho2k}) are feasible for the definitions
in (\ref{eq:genie1}) and (\ref{eq:genie2}) if and only if $a_j\leq
d_j$ and $b_k\leq c_k$. Moreover, (\ref{eq:s1i})-(\ref{eq:rho2i}),
(\ref{eq:s1j}) and (\ref{eq:s2k}) satisfy
\bqa%
\sqrt{c_l}\rho_{1l}\sigma_{1l}&{}={}&1+a_lQ_l^*,\label{eq:rho1isigma1i}\\
\sqrt{d_l}\rho_{2l}\sigma_{2l}&{}={}&1+b_lP_l^*,\label{eq:rho2isigma2i}%
\eqa%
for all $l=1,\cdots,m$.

Let $\epsilon>0$ and $\epsilon\rightarrow 0$ as
$n\rightarrow\infty$. From Fano's inequality, any achievable rate
$R_1$ and $R_2$ for the PGIC must satisfy
\bqa%
&&n(R_1+R_2)-n\epsilon\nn\\
&&\leq I\left(\xp_1^n;\yp_1^n\right)+
I\left(\xp_2^n;\yp_2^n\right)\nn\\
&&\leq
I\left(\xp_1^n;\yp_1^n,\xp_{1\mi}^n+\np_{1\mi}^n,\xp_{1\mj}^n+\np_{1\mj}^n\right)+
I\left(\xp_2^n;\yp_2^n,\xp_{2\mi}^n+\np_{2\mi}^n,\xp_{2\mk}^n+\np_{2\mk}^n\right)\nn\\
&&=h\left(\yp_1^n,\xp_{1\mi}^n+\np_{1\mi}^n,\xp_{1\mj}^n+\np_{1\mj}^n\right)
-h\left(\yp_1^n,\xp_{1\mi}^n+\np_{1\mi}^n,\xp_{1\mj}^n+\np_{1\mj}^n\left|\hspace{.03in}\xp_1^n\right.\right)
+h\left(\yp_2^n,\xp_{2\mi}^n+\np_{2\mi}^n,\xp_{2\mk}^n+\np_{2\mk}^n\right)\nn\\
&&\hspace{.2in}-h\left(\yp_2^n,\xp_{2\mi}^n+\np_{2\mi}^n,\xp_{2\mk}^n+\np_{2\mk}^n\left|\hspace{.03in}\xp_2^n\right.\right).
\label{eq:Fano}%
\eqa
In (\ref{eq:Fano}), we provide side information
$\xp_{1\mi}^n+\np_{1\mi}^n$ and $\xp_{1\mj}^n+\np_{1\mj}^n$ to
receiver one, and $\xp_{2\mi}^n+\np_{2\mi}^n$ and
$\xp_{2\mk}^n+\np_{2\mk}^n$ to receiver two, respectively. For the
first $m_1$ sub-channels which have two-sided interference, both
receivers have side information. For the sub-channels which have
one-sided interference, only the receivers suffering from
interference have the corresponding side information. For the
sub-channels without interference, no side information is given.

For the first term of (\ref{eq:Fano}), we have
\bqa%
&&h\left(\yp_1^n,\xp_{1\mi}^n+\np_{1\mi}^n,\xp_{1\mj}^n+\np_{1\mj}^n\right)\nn\\
&&=h\left(\yp_{1\mi}^n,\yp_{1\mj}^n,\yp_{1\mk}^n,\yp_{1\mr}^n,\xp_{1\mi}^n+\np_{1\mi}^n,\xp_{1\mj}^n+\np_{1\mj}^n\right)\nn\\
&&\leq h\left(\yp_{1\mk}^n,\xp_{1\mi}^n+\np_{1\mi}^n\right)
+h\left(\yp_{1\mi}^n\left|\hspace{.03in}\xp_{1\mi}^n+\np_{1\mi}^n\right.\right)
+h\left(\yp_{1\mj}^n,\xp_{1\mj}^n+\np_{1\mj}^n\right)+h\left(\yp_{1\mr}^n\right)\nn\\
&&\leq h\left(\yp_{1\mk}^n,\xp_{1\mi}^n+\np_{1\mi}^n\right)
+nh\left(\widehat\yp_{1\mi}^*\left|\hspace{.03in}\widehat\xp_{1\mi}^*+\np_{1\mi}\right.\right)
+nh\left(\widehat\yp_{1\mj}^*,\widehat\xp_{1\mj}^*+\np_{1\mj}\right)+nh\left(\widehat\yp_{1\mr}^*\right)\nn\\
&&\leq
h\left(\yp_{1\mk}^n,\xp_{1\mi}^n+\np_{1\mi}^n\right)+n\sum_ih\left(\widehat
Y_{1i}^*\left|\hspace{.03in}\widehat X_{1i}^*+N_{1i}\right.\right)
+n\sum_jh\left(\widehat Y_{1j}^*,\widehat
X_{1j}^*+N_{1j}\right)+n\sum_rh\left(\widehat Y_{1r}^*\right),\label{eq:first}%
\eqa%
where the first inequality follows by the chain rule and the fact
that conditioning does not increase entropy, and the second
inequality is from Lemma \ref{lemma:generalconcave}.

For the fourth term of (\ref{eq:Fano}), we have
\bqa%
&&-h\left(\yp_2^n,\xp_{2\mi}^n+\np_{2\mi}^n,\xp_{2\mk}^n+\np_{2\mk}^n\left|\hspace{.03in}\xp_2^n\right.\right)\nn\\
&&=-h\left(\yp_{2\mi}^n,\yp_{2\mj}^n,\yp_{2\mk}^n,\yp_{2\mr}^n,\xp_{2\mi}^n+\np_{2\mi}^n,\xp_{2\mk}^n+\np_{2\mk}^n\left|\hspace{.03in}\xp_{2\mi}^n,\xp_{2\mj}^n,\xp_{2\mk}^n,\xp_{2\mr}^n\right.\right)\nn\\
&&=-h\left(\Bbf_\mi\xp_{1\mi}^n+\zp_{2\mi}^n,\zp_{2\mj}^n,\Bbf_\mk\xp_{1\mk}^n+\zp_{2\mk}^n,\zp_{2\mr}^n,\np_{2\mi}^n,\np_{2\mk}^n\right)\nn\\
&&=-h\left(\Bbf_\mi\xp_{1\mi}^n+\zp_{2\mi}^n,\Bbf_\mk\xp_{1\mk}^n+\zp_{2\mk}^n,\np_{2\mi}^n,\np_{2\mk}^n\right)-h\left(\zp_{2\mj}^n\right)-h\left(\zp_{2\mr}^n\right)\nn\\
&&=-h\left(\Bbf_\mi\xp_{1\mi}^n+\zp_{2\mi}^n,\Bbf_\mk\xp_{1\mk}^n+\zp_{2\mk}^n\left|\hspace{.03in}\np_{2\mi}^n,\np_{2\mk}^n\right.\right)
-h\left(\np_{2\mi}^n\right)-h\left(\np_{2\mk}^n\right)-h\left(\zp_{2\mj}^n\right)-h\left(\zp_{2\mr}^n\right)\nn\\
&&=-h\left(\Bbf_\mi\xp_{1\mi}^n+\zp_{2\mi}^n,\Bbf_\mk\xp_{1\mk}^n+\zp_{2\mk}^n\left|\hspace{.03in}\np_{2\mi}^n,\np_{2\mk}^n\right.\right)
-n\sum_ih\left(N_{2i}\right)-n\sum_kh\left(N_{2k}\right)-n\sum_jh\left(Z_{2j}\right)\nn\\
&&\hspace{.2in}-n\sum_rh\left(Z_{2r}\right).\label{eq:fourth}
\eqa%
 where the third equality holds since $\zp_{2\mj}^n$ and
 $\zp_{2\mr}^n$ are
independent of all other variables.

Combine the first terms of (\ref{eq:first}) and (\ref{eq:fourth}),
we have
\bqa &&h\left(\yp_{1\mk}^n,\xp_{1\mi}^n+\np_{1\mi}^n\right)
-h\left(\Bbf_\mi\xp_{1\mi}^n+\zp_{2\mi}^n,\Bbf_\mk\xp_{1\mk}^n+\zp_{2\mk}^n\left|\hspace{.03in}\np_{2\mi}^n,\np_{2\mk}^n\right.\right)\nn\\
&&\stackrel{(a)}=h\left(\xp_{1\mi}^n+\np_{1\mi}^n,\Cbf_\mk\xp_{1\mk}^n+\zp_{1\mk}^n\right)
-h\left(\Bbf_\mi\xp_{1\mi}^n+\wp_{2\mi}^n,\Bbf_\mk\xp_{1\mk}^n+\wp_{2\mk}^n\right)\nn\\
&&=h\left(\xp_{1\mi}^n+\np_{1\mi}^n,\xp_{1\mk}^n+\Cbf_\mk^{-1}\zp_{1\mk}^n\right)
-h\left(\xp_{1\mi}^n+\Bbf_\mi^{-1}\wp_{2\mi}^n,\xp_{1\mk}^n+\Bbf_\mk^{-1}\wp_{2\mk}^n\right)+n\log\frac{|\Cbf_\mk|}{|\Bbf_\mi|\cdot|\Bbf_\mk|}\nn\\
&&\stackrel{(b)}=n\log\frac{|\Cbf_\mk|}{|\Bbf_\mi|\cdot|\Bbf_\mk|}\nn\\
&&\stackrel{(c)}=n\sum_ih\left(\widehat
X_{1i}^*+N_{1i}\right)+n\sum_kh\left(\widehat
X_{1k}^*+\frac{1}{\sqrt{c_k}}Z_{1k}\right)
-n\sum_ih\left(\widehat X_{1i}^*+\frac{1}{\sqrt{b_i}}W_{2i}\right)\nn\\
&&\hspace{.2in}-n\sum_kh\left(\widehat X_{1k}^*+\frac{1}{\sqrt{b_k}}W_{2k}\right)+n\sum_k\log\sqrt{c_k}-n\sum_i\log\sqrt{b_i}-n\sum_k\log\sqrt{b_k}\nn\\
&&=n\sum_ih\left(\widehat
X_{1i}^*+N_{1i}\right)+n\sum_kh\left(\sqrt{c_k}\widehat
X_{1k}^*+Z_{1k}\right)
-n\sum_ih\left(\sqrt{b_i}\widehat X_{1i}^*+Z_{2i}\left|\hspace{.02in}N_{2i}\right.\right)\nn\\
&&\hspace{.2in}-n\sum_kh\left(\sqrt{b_k}\widehat
X_{1k}^*+Z_{2k}\left|\hspace{.02in}N_{2k}\right.\right),%
\label{eq:cancel}%
\eqa
 where in (a) we let $\wp_{2\mi}$ and $\wp_{2\mk}$ be
independent Gaussian vectors and\bqa
&&\Cov\left(\wp_{2\mi}\right)=\Cov\left(\zp_{2\mi}\left|\hspace{.02in}\np_{2\mi}\right.\right)=\Ibf_\mi-\diag(\rho_{2\mi}^2)\nn\\
&&\Cov\left(\wp_{2\mk}\right)=\Cov\left(\zp_{2\mk}\left|\hspace{.02in}\np_{2\mk}\right.\right)=\Ibf_\mk-\diag(\rho_{2\mk}^2).
 \eqa
The stacked vectors $\wp_{2\mi}^n$ and $\wp_{2\mk}^n$ each have
independent and identical distribution (i.i.d) entries. Equality
(b) holds because of (\ref{eq:rho2i}) and (\ref{eq:rho2k}) which
imply \bqn
\Cov\left(\np_{1\mi}\right)&{}={}&\Cov\left(\Bbf_\mi^{-1}\wp_{2\mi}\right),\\
\Cov\left(\Cbf_\mk^{-1}\zp_{1\mk}\right)&{}={}&\Cov\left(\Bbf_\mk^{-1}\wp_{2\mk}\right),\eqn
and \bqn
h\left(\xp_{1\mi}^n+\np_{1\mi}^n,\xp_{1\mk}^n+\Cbf_\mk^{-1}\zp_{1\mk}^n\right)
-h\left(\xp_{1\mi}^n+\Bbf_\mi^{-1}\wp_{2\mi}^n,\xp_{1\mk}^n+\Bbf_\mk^{-1}\wp_{2\mk}^n\right)=0,\eqn
regardless of the distribution of $\xp_{1\mi}^n$ and
$\xp_{1\mk}^n$.

Equality (c) holds also because of (\ref{eq:rho2i}) and
(\ref{eq:rho2k}), which imply %
\bqn%
 h\left(\widehat
X_{1i}^*+N_{1i}\right)-h\left(X_{1i}^n+\frac{1}{\sqrt{b_i}}W_{2i}\right)&{}={}&0,\\
h\left(\widehat
X_{1k}^*+\frac{1}{\sqrt{c_k}}Z_{1k}\right)-h\left(\widehat
X_{1k}^*+\frac{1}{\sqrt{b_k}}W_{2k}\right)&{}={}&0,%
\eqn%
 regardless of the distributions of $\widehat X_{1i}$ and
$\widehat X_{1k}$.

Combining (\ref{eq:first}) and (\ref{eq:fourth}) and using
(\ref{eq:cancel}), we have%
\bqa%
&&h\left(\yp_1^n,\xp_{1\mi}^n+\np_{1\mi}^n,\xp_{1\mj}^n+\np_{1\mj}^n\right)
-h\left(\yp_2^n,\xp_{2\mi}^n+\np_{2\mi}^n,\xp_{2\mk}^n+\np_{2\mk}^n\left|\hspace{.03in}\xp_2^n\right.\right)\nn\\
&&\leq n\sum_ih\left(\widehat
X_{1i}^*+N_{1i}\right)+n\sum_kh\left(\sqrt{c_k}\widehat
X_{1k}^*+Z_{1k}\right)+n\sum_ih\left(\widehat
Y_{1i}^*\left|\hspace{.03in}\widehat
X_{1i}^*+N_{1i}\right.\right)+n\sum_rh\left(\widehat
Y_{1r}^*\right)
\nn\\
&&\hspace{.2in}+n\sum_jh\left(\widehat Y_{1j}^*,\widehat
X_{1j}^*+N_{1j}\right)-n\sum_ih\left(\sqrt{b_i}\widehat
X_{1i}^*+Z_{2i}\left|\hspace{.02in}N_{2i}\right.\right)-n\sum_kh\left(\sqrt{b_k}\widehat
X_{1k}^*+Z_{2k}\left|\hspace{.02in}N_{2k}\right.\right)\nn\\
&&\hspace{.2in}-n\sum_ih\left(N_{2i}\right)-n\sum_kh\left(N_{2k}\right)-n\sum_jh\left(Z_{2j}\right)-n\sum_rh\left(Z_{2r}\right).%
\label{eq:half1}%
\eqa%

Similarly, because of (\ref{eq:rho1i}) and (\ref{eq:rho1j}) we
have%
\bqa%
&&h\left(\yp_2^n,\xp_{2\mi}^n+\np_{2\mi}^n,\xp_{2\mk}^n+\np_{2\mk}^n\right)
 -h\left(\yp_1^n,\xp_{1\mi}^n+\np_{1\mi}^n,\xp_{1\mj}^n+\np_{1\mj}^n\left|\hspace{.03in}\xp_1^n\right.\right)
 \nn\\
 &&\leq n\sum_ih\left(\widehat
 X_{2i}^*+N_{2i}\right)+n\sum_jh\left(\sqrt{d_j}\widehat
 X_{2j}^*+Z_{2j}\right)+n\sum_i\left(Y_{2i}^*\left|\hspace{.02in}\widehat
 X_{2i}^*+N_{2i}\right.\right)+n\sum_rh\left(\widehat Y_{2r}^*\right)\nn\\
 &&\hspace{.2in}+n\sum_kh\left(\widehat Y_{2k}^*,\widehat
 X_{2k}^*+N_{2k}\right)
 -n\sum_ih\left(\sqrt{a_i}\widehat X_{2i}^*+Z_{1i}\left|\hspace{.02in}N_{1i}\right.\right)
 -n\sum_jh\left(\sqrt{a_j}\widehat
 X_{2j}^*+Z_{1j}\left|\hspace{.02in}N_{1j}\right.\right)\nn\\
 &&\hspace{.2in}-n\sum_ih\left(N_{1i}\right)-n\sum_jh\left(N_{1j}\right)-n\sum_kh\left(Z_{1k}\right)-n\sum_rh\left(Z_{1r}\right).%
\label{eq:half2}%
\eqa%
Substituting (\ref{eq:half1}) and
(\ref{eq:half2}) into (\ref{eq:Fano}), we have
\bqa%
&& R_1+R_2-\epsilon\nn\\
&&\leq \sum_i\left[h\left(\widehat
X_{1i}^*+N_{1i}\right)+h\left(\widehat
Y_{1i}^*\left|\hspace{.02in}\widehat
X_{1i}^*+N_{1i}\right.\right)-h\left(N_{1i}\right)-h\left(\sqrt{a_i}\widehat
X_{2i}^*+Z_{1i}\left|\hspace{.02in}N_{1i}\right.\right)\right.\nn\\
&&\hspace{.6in}\left.h\left(\widehat
X_{2i}^*+N_{2i}\right)+h\left(\widehat
Y_{2i}^*\left|\hspace{.02in}\widehat
X_{2i}^*+N_{2i}\right.\right)-h\left(N_{2i}\right)-h\left(\sqrt{b_i}\widehat
X_{1i}^*+Z_{2i}\left|\hspace{.02in}N_{2i}\right.\right)\right]\nn\\
&&\hspace{.2in}+\sum_j\left[h\left(\widehat Y_{1j}^*,\widehat
X_{1j}^*+N_{1j}\right)-h\left(N_{1j}\right)-h\left(\sqrt{a_j}\widehat
X_{2j}^*+Z_{1j}\left|\hspace{.02in}N_{1j}\right.\right)+h\left(\sqrt{d_j}\widehat
X_{2j}^*+Z_{2j}\right)-h\left(Z_{2j}\right)\right]\nn\\
&&\hspace{.2in}+\sum_k\left[h\left(\widehat Y_{2k}^*,\widehat
X_{2k}^*+N_{2k}\right)-h\left(N_{2k}\right)-h\left(\sqrt{b_k}\widehat
X_{1k}^*+Z_{2k}\left|\hspace{.02in}N_{2k}\right.\right)+h\left(\sqrt{c_k}\widehat
X_{1k}^*+Z_{1k}\right)-h\left(Z_{1k}\right)\right]\nn\\
&&\hspace{.2in}+\sum_r\left[h\left(\widehat Y_{1r}^*\right)-h\left(Z_{1r}\right)\right]%
   +\sum_r\left[h\left(\widehat
   Y_{2r}^*\right)-h\left(Z_{2r}\right)\right]\nn\\
&&=\sum_if_i(P_i,Q_i)+\sum_jf_j(P_j,Q_j)+\sum_kf_k(P_k,Q_k)
+\sum_rf_r(P_r,Q_r)\nn\\
&&=\sum_{l=1}^mf_l(P_l,Q_l), %
\label{eq:funcSum}%
\eqa
where
\bqa%
&&\hspace{-.2in}f_i(P_i,Q_i)=\frac{1}{2}\log\left[\frac{\left(1+a_iQ_{i}\right)P_{i}}{1+a_iQ_{i}-\rho_{1i}^2}
\left(\frac{1}{\sigma_{1i}}-\frac{\sqrt{c_i}\rho_{1i}}{1+a_iQ_{i}}\right)^2+1+\frac{c_iP_{i}}{1+a_iQ_{i}}\right]\nn\\
&&\hspace{.65in}+\frac{1}{2}\log\left[\frac{\left(1+b_iP_{i}\right)Q_{i}}{1+b_iP_{i}-\rho_{2i}^2}
\left(\frac{1}{\sigma_{2i}}-\frac{\sqrt{d_i}\rho_{2i}}{1+b_iP_{i}}\right)^2+1+\frac{d_iQ_{i}}{1+b_iP_{i}}\right],\label{eq:fi}\\
&&\hspace{-.2in}f_j(P_j,Q_j)=\frac{1}{2}\log\left[\frac{\left(1+a_jQ_{j}\right)P_{j}}{1+a_jQ_{j}-\rho_{1j}^2}
\left(\frac{1}{\sigma_{1j}}-\frac{\sqrt{c_j}\rho_{1j}}{1+a_jQ_{j}}\right)^2+1+\frac{c_jP_{j}}{1+a_jQ_{j}}\right]+\frac{1}{2}\log\left(1+d_jQ_{j}\right),\label{eq:fj}\\
&&\hspace{-.2in}f_k(P_k,Q_k)=\frac{1}{2}\log\left[\frac{\left(1+b_kP_{k}\right)Q_{k}}{1+b_kP_{k}-\rho_{2k}^2}
\left(\frac{1}{\sigma_{2k}}-\frac{\sqrt{d_k}\rho_{2k}}{1+b_kP_{k}}\right)^2+1+\frac{d_kQ_{k}}{1+b_kP_{k}}\right]+\frac{1}{2}\log\left(1+c_kP_{k}\right),\label{eq:fk}\\
&&\hspace{-.2in}f_r(P_r,Q_r)=\frac{1}{2}\log\left(1+P_{r}\right)+\frac{1}{2}\log\left(1+Q_{r}\right).\label{eq:fr}%
\eqa
Next we will show that the $f_l(P_l,Q_l)$, $l=1,\cdots,m$ are all
concave and non-decreasing functions of $(P_l,Q_l)$. From
(\ref{eq:funcSum}) and the fact that
\bqn%
&&h\left(\widehat X_{1i}^*+N_{1i}\right)-h\left(\sqrt{b_i}\widehat
X_{1i}^*+Z_{2i}\left|\hspace{.02in}N_{2i}\right.\right)=-\log\sqrt{b_i},\\
&&h\left(\widehat X_{2i}^*+N_{2i}\right)-h\left(\sqrt{a_i}\widehat
X_{2i}^*+Z_{1i}\left|\hspace{.02in}N_{1i}\right.\right)=-\log\sqrt{a_i},%
\eqn
we have
\bqa%
f_i(P_i,Q_i)=h\left(\widehat Y_{1i}^*\left|\hspace{.02in}\widehat
X_{1i}^*+N_{1i}\right.\right)-h\left(N_{1i}\right)+h\left(\widehat
Y_{2i}^*\left|\hspace{.02in}\widehat
X_{2i}^*+N_{2i}\right.\right)-h\left(N_{2i}\right)-\log\sqrt{a_ib_i}.%
\eqa
Define Gaussian variables $\widehat X_{1i\_t}^*$, $\widehat
X_{2i\_t}^*$ and $\widehat Y_{1i\_t}^*$ independent of $N_i$, and
$\widehat Y_{1i\_t}^*=\sqrt{c_i}\widehat
X_{1i\_t}^*+\sqrt{a_i}\widehat X_{2i\_t}^*+Z_i$, $t=1,\cdots,s$
where $s$ is an integer. Let $\Var\left(\widehat
X_{1i\_t}^*\right)=P_{i\_t}$, $\Var\left(\widehat
X_{2i\_t}^*\right)=Q_{i\_t}$ and
\bqa%
&&\sum_{t=1}^s\lambda_t\Var\left(\widehat
X_{1i\_t}^*\right)=P_i=\Var\left(\widehat
X_{1i}^*\right),\\
&&\sum_{t=1}^s\lambda_t\Var\left(\widehat
X_{2i\_t}^*\right)=Q_i=\Var\left(\widehat
X_{2i}^*\right),%
\eqa
where $\{\lambda_t\}$ is a non-negative sequence with
$\sum_{t=1}^s\lambda_t=1$. Then we have
\bqa \sum_{t=1}^s\lambda_l\Cov\left(\left[\begin{array}{c}
  \widehat Y_{1i\_t}^* \\
  \widehat X_{1i\_t}^*+N_i \\
\end{array}\right]\right)=\Cov\left(\left[\begin{array}{c}
  \widehat Y_{1i}^* \\
  \widehat X_{1i}^*+N_i \\
\end{array}\right]\right),%
\eqa
From Lemma \ref{lemma:generalconcave} we have
\bqa%
 h\left(\widehat
Y_{1i}^*\left|\hspace{.02in}\widehat
X_{1i}^*+N_{1i}\right.\right)\geq\sum_{l=1}^k\lambda_th\left(\widehat
Y_{1i\_t}^*\left|\hspace{.02in}\widehat
X_{1i\_t}^*+N_{1i}\right.\right).%
\eqa
Therefore, $h\left(\widehat Y_{1i}^*\left|\hspace{.02in}\widehat
X_{1i}^*+N_{1i}\right.\right)$ is a concave function of
$(P_i,Q_i)$. For the same reason $h\left(\widehat
Y_{2i}^*\left|\hspace{.02in}\widehat
X_{2i}^*+N_{2i}\right.\right)$ is also a concave function of
$(P_i,Q_i)$. Therefore $f_i(P_i,Q_i)$ is a concave function of
$(P_i,Q_i)$. Similar steps show that $f_j$, $f_k$, and $f_r$ are
concave in $(P_l,Q_l)$.

To show that $f_i(P_i,Q_i)$ is a non-decreasing function of
$P_i,Q_i$, we let $\bar X_{1i}$, $\bar X_{2i}$, $U_1$ and $U_2$ be
four independent Gaussian variables and
\bqn%
&&\widehat X_{1i}^*=\bar X_{1i}+U_1,\\
&&\widehat X_{2i}^*=\bar X_{2i}+U_2.%
 \eqn
Then
\bqn%
&&h\left(\widehat Y_{1i}^*\left|\widehat
X_{1i}^*+N_{1i}\right.\right)\\
&&=h\left(\sqrt{c_i}\widehat X_{1i}^*+\sqrt{a_i}\widehat
X_{2i}^*+Z_{1i}\left|\widehat
X_{1i}^*+N_{1i}\right.\right)\\
&&\leq h\left(\sqrt{c_i}\widehat X_{1i}^*+\sqrt{a_i}\widehat
X_{2i}^*+Z_{1i}\left|\widehat
X_{1i}^*+N_{1i},U_1,U_2\right.\right)\\
&&=h\left(\sqrt{c_i}\bar X_{1i}+\sqrt{a_i}\bar
X_{2i}+Z_{1i}\left|\bar X_{1i}+N_{1i}\right.\right).%
\eqn
Therefore, $h\left(\widehat Y_{1i}^*\left|\widehat
X_{1i}^*+N_{1i}\right.\right)$ is a non-decreasing function of
$(P_i,Q_i)$. For the same reason, $h\left(\widehat
Y_{2i}^*\left|\widehat X_{2i}^*+N_{2i}\right.\right)$ is also a
non-decreasing function of $(P_i,Q_i)$. Therefore, $f_i$ is a
non-decreasing function of $(P_i,Q_i)$.

From (\ref{eq:rho1isigma1i}) and (\ref{eq:rho2isigma2i}) we have
\bqa
\sum_{l=1}^mf_l(P_l^*,Q_l^*)&{}={}&\frac{1}{2}\sum_{l=1}^m\left[\log\left(1+\frac{c_lP_l^*}{1+a_lQ_l^*}\right)+\log\left(1+\frac{d_lQ_l^*}{1+b_lP_l^*}\right)\right]\nn\\
&{}={}&\sum_{l=1}^mC_l(P_l^*,Q_l^*).\label{eq:mathfunc}\eqa Next
we will show that $\sum_{l=1}^mf(P_l^*,Q_l^*)\geq
\sum_{l=1}^mf(P_l,Q_l)$ for any $P_l,Q_l$ that satisfy
$\sum_{l=1}^mP_l=P$, $\sum_{l=1}^mQ_l=Q$.

Using (\ref{eq:rho1isigma1i}) and (\ref{eq:rho2isigma2i}), we have
\bqa%
&&\frac{\partial f_l(p,q)}{\partial p}\left|_{\tiny
\begin{array}{c}
  p=P_l^* \\
  q=Q_l^* \\
\end{array}}\right.=\frac{\partial C_l(p,q)}{\partial p}\left|_{\tiny\begin{array}{c}
  p=P_l^* \\
  q=Q_l^* \\
\end{array}}\right.\\
&&\frac{\partial f_l(p,q)}{\partial
q}\left|_{\tiny\begin{array}{c}
  p=P_l^* \\
  q=Q_l^* \\
\end{array}}\right.=\frac{\partial C_l(p,q)}{\partial q}\left|_{\tiny \begin{array}{c}
  p=P_l^* \\
  q=Q_l^* \\
\end{array}}\right.,%
\eqa
for all $l=1,\cdots,m$. Therefore, $f_l$ and $C_l$ have the same
partial derivatives at point $(P_l^*,Q_l^*)$. From the Appendix,
the subgradient of a function is determined by the derivatives,
therefore, $f_l$ and $C_l$ have the same subgradient at point
$(P_l^*,Q_l^*)$ for each $l$. From (\ref{eq:identicalsg}), we have
\bqa%
\left[k_p^*,k_q^*\right]\in\partial f_l(P_l^*,Q_l^*),\quad
l=1,\cdots,m. \eqa
Therefore, from Lemma \ref{lemma:PGIC_subg} we have
\bqa%
\sum_{l=1}^mf_l\left(P_l^*,Q_l^*\right)\geq\sum_{l=1}^mf_l\left(P_l,Q_l\right),%
\eqa
if $\sum_{l=1}^mP_l=P$ and $\sum_{l=1}^mQ_l=Q$.

If $\sum_{l=1}^mP_l\leq P$ and $\sum_{l=1}^mQ_l\leq Q$, there
exist two non-negative sequences $\{s_1,\cdots,s_m\}$ and
$\{t_1,\cdots,t_m\}$ such that $\sum_{l=1}^m(P_l+s_l)=P$ and
$\sum_{l=1}^m(Q_l+t_l)=Q$. Therefore we have from Lemma
\ref{lemma:PGIC_subg}
\bqa%
\sum_{l=1}^mf_l\left(P_l^*,Q_l^*\right)\geq\sum_{i=1}^mf_l\left(P_l+s_l,Q_l+t_l\right).%
\label{eq:step1}%
\eqa
Since $f_l(P_l,Q_l)$, $l=1,\cdots,m$ are all non-decreasing
functions of $P_l$ and $Q_l$, we have
\bqa%
\sum_{i=1}^mf_l\left(P_l+s_l,Q_l+t_l\right)\geq\sum_{i=1}^mf_l\left(P_l,Q_l\right).%
\label{eq:step2}%
\eqa
Combining (\ref{eq:step1}) and (\ref{eq:step2}) we have \bqa
\sum_{l=1}^mf_l\left(P_l^*,Q_l^*\right)\geq\sum_{i=1}^mf_l\left(P_l,Q_l\right),\eqa
for any $\sum_{l=1}^mP_l\leq P$ and $\sum_{l=1}^mQ_l\leq Q$.
Therefore we have from (\ref{eq:funcSum}) and (\ref{eq:mathfunc}) that
\bqa%
R_1+R_2-\epsilon\leq\frac{1}{2}\sum_{l=1}^m\left[\log\left(1+\frac{c_lP_l^*}{1+a_lQ_l^*}\right)+\log\left(1+\frac{d_lQ_l^*}{1+b_lP_l^*}\right)\right].%
\label{eq:converse}%
\eqa
The above sum rate is achievable by independent transmission
across sub-channels and single-user detection in each sub-channel.
Therefore (\ref{eq:converse}) is the sum-rate capacity of the PGIC
if the power constraints satisfy (\ref{eq:PGIC_generalPower}).
\epf

\begin{figure}[htp]
\centerline{\leavevmode \epsfxsize=5in \epsfysize=4in
\epsfbox{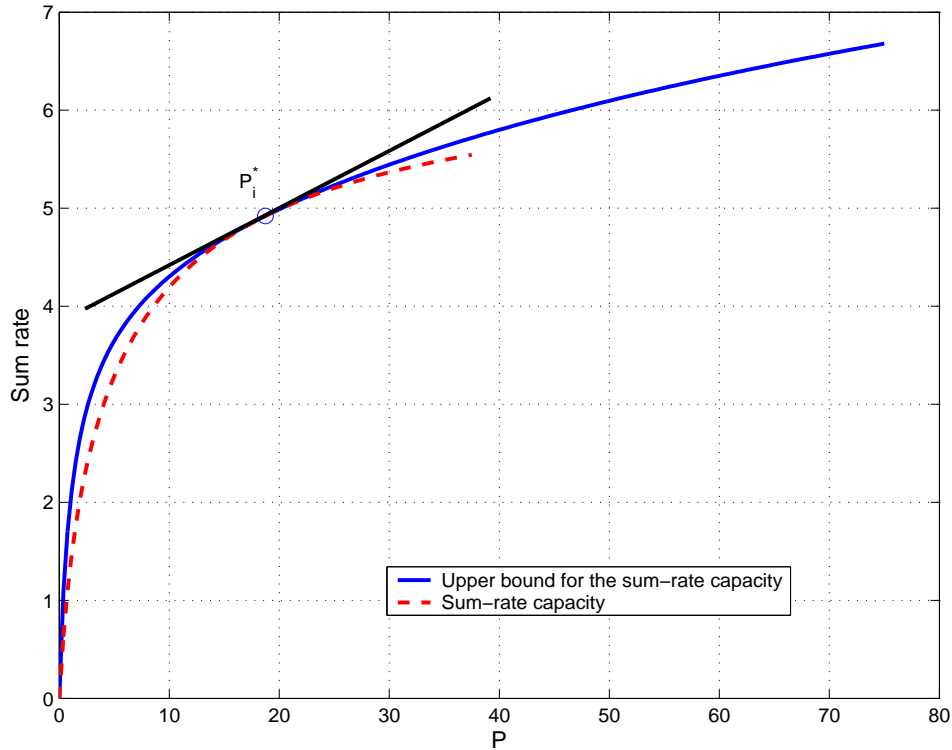}}\caption{An illustration of the upper
bound suitable for the proof of Theorem \ref{thm:pgic}.}
\label{fig:PGIC_goodbound}\end{figure}

{\em Remark 7: } The main idea of the proof can be summarized as
follows. We first assume an arbitrary power allocation
$(P_l,Q_l)$, $l=1,\cdots,m$. Then we show that the sum rate for
this power allocation is upper bounded by
$\sum_{l=1}^mf_l\left(P_l,Q_l\right)$. This upper bound decomposes
the sum rate bound into the sum of the individual sub-channel's
sum-rate capacity upper bounds. By Lemma \ref{lemma:PGIC_subg},
the maximum of $\sum_{l=1}^mf_l\left(P_l,Q_l\right)$ is
$\sum_{l=1}^mC_l\left(P_l^*,Q_l^*\right)$ which is an achievable
sum rate for a special power allocation. To ease the proof, the
upper bound $f_l$ can not be arbitrarily chosen. Compared to the
sum-rate capacity $C_l$ for sub-channel $l$, $f_l$ has the
following properties:
\bi%
\item $f_l$ is concave over the powers; %
\item $f_l$ is tight at the
optimal point $(P_l^*,Q_l^*)$;%
\item $f_l$ and $C_l$ have the same subdifferentials
at the
optimal point $(P_l^*,Q_l^*)$.%
\ei
Therefore we choose the noise vectors $\np_1$ and $\np_2$ such
that the above conditions are satisfied. Fig.
\ref{fig:PGIC_goodbound} illustrates such an upper bound.

\section{Numerical examples}%
 \label{section:numerical}

\begin{figure}[htp]
\centerline{\leavevmode \epsfxsize=5.5in \epsfysize=4.5in
\epsfbox{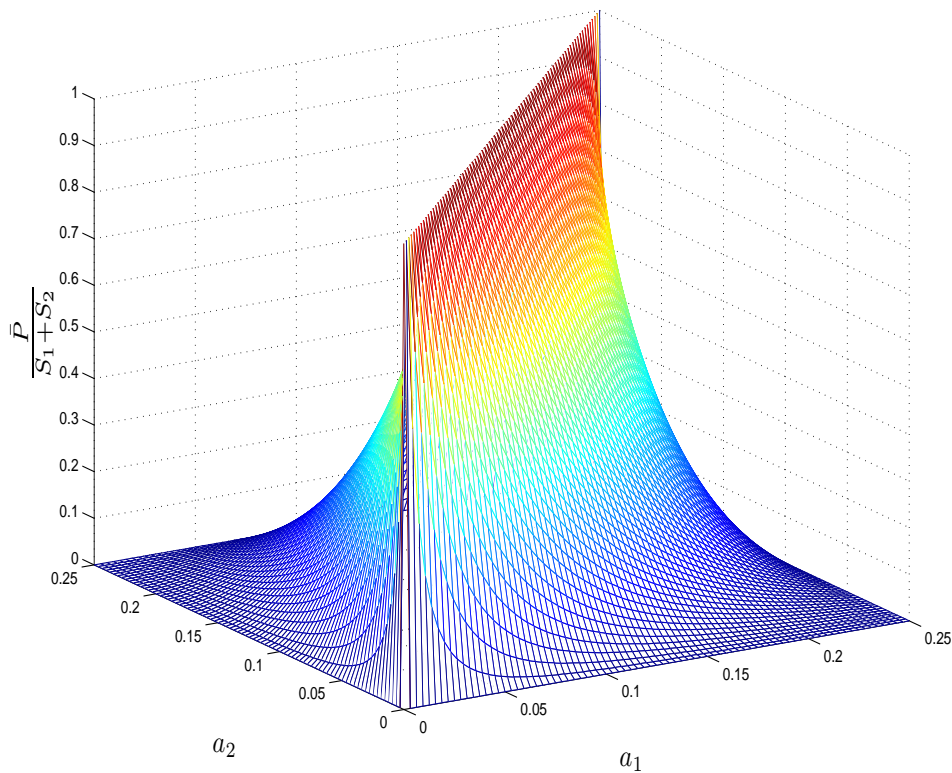}}\caption{Ratio of the total power
constraint and the sum of the individual sub-channel power
constraints for different channel gains.}
\label{fig:PGIC_ratio}\end{figure}

\begin{figure}[htp]
\centerline{\leavevmode \epsfxsize=5.5in \epsfysize=4.5in
\epsfbox{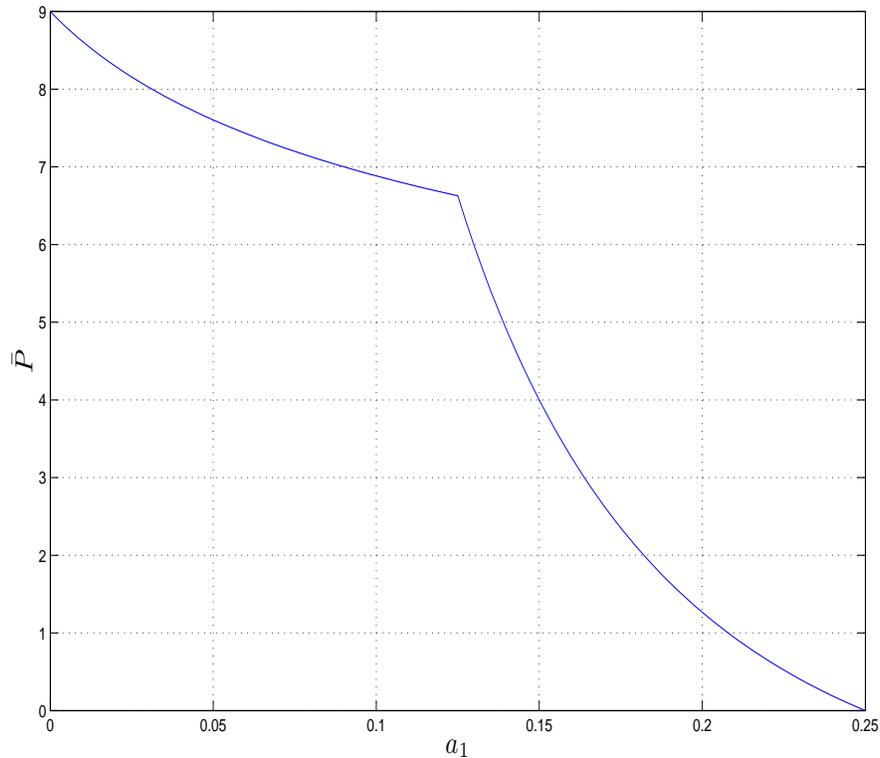}}\caption{The maximum noisy interference power
constraint $\bar P$ for $a_1$ with $a_2=\frac{1}{8}$. }
\label{fig:PGIC_Pbar}\end{figure}

Figs. \ref{fig:PGIC_ratio} and \ref{fig:PGIC_Pbar} are examples
for a symmetric PGIC with directly link gain $c_i=d_i=1$. Fig.
\ref{fig:PGIC_ratio} shows the ratio of maximum noisy interference
power constraint $\bar P$ for a two-channel symmetric PGIC with
$a$ varying from $0$ to $0.25$, and the sum of the maximum noisy
interference power constraints for
 $S_1+S_2$ where $S_i=\frac{\sqrt{a_i}-2a_i}{2a_i^2}$, $i=1,2$. Thus, for $a_1=a_2$ (i.e., these two sub-channels are identical),
   the ratio is $1$ and $S_1=S_2=\frac{\bar
 P}{2}$, and when $a_1$ and $a_2$ are far apart, then $\bar P\ll
 S_1+S_2$. Thus, for $a_1=a_2$
we achieve capacity despite the fact that we do not know the
capacity if all power is placed in one sub-channel. Again, this is
where Lemma \ref{lemma:PGIC_sumConcave} is useful.

 Fig. \ref{fig:PGIC_Pbar} shows the maximum noisy interference
power constraint $\bar P$ for a two-channel symmetric PGIC with
$a_1$ varying in $[0,\frac{1}{4}]$ and $a_2=\frac{1}{8}$. When
$a_1=\frac{1}{4}$, the first sub-channel no longer has noisy
interference, therefore the maximum noisy interference power
constraint is $\bar P=0$. Fig. \ref{fig:PGIC_Pbar} also shows that
$\bar P$ decreases with $a_1$. The discontinuity at $a_1=
\frac{1}{8}$ is because the second sub-channel becomes the worse
channel.

\begin{figure}[htp]
\centerline{\leavevmode \epsfxsize=5.5in \epsfysize=4.5in
\epsfbox{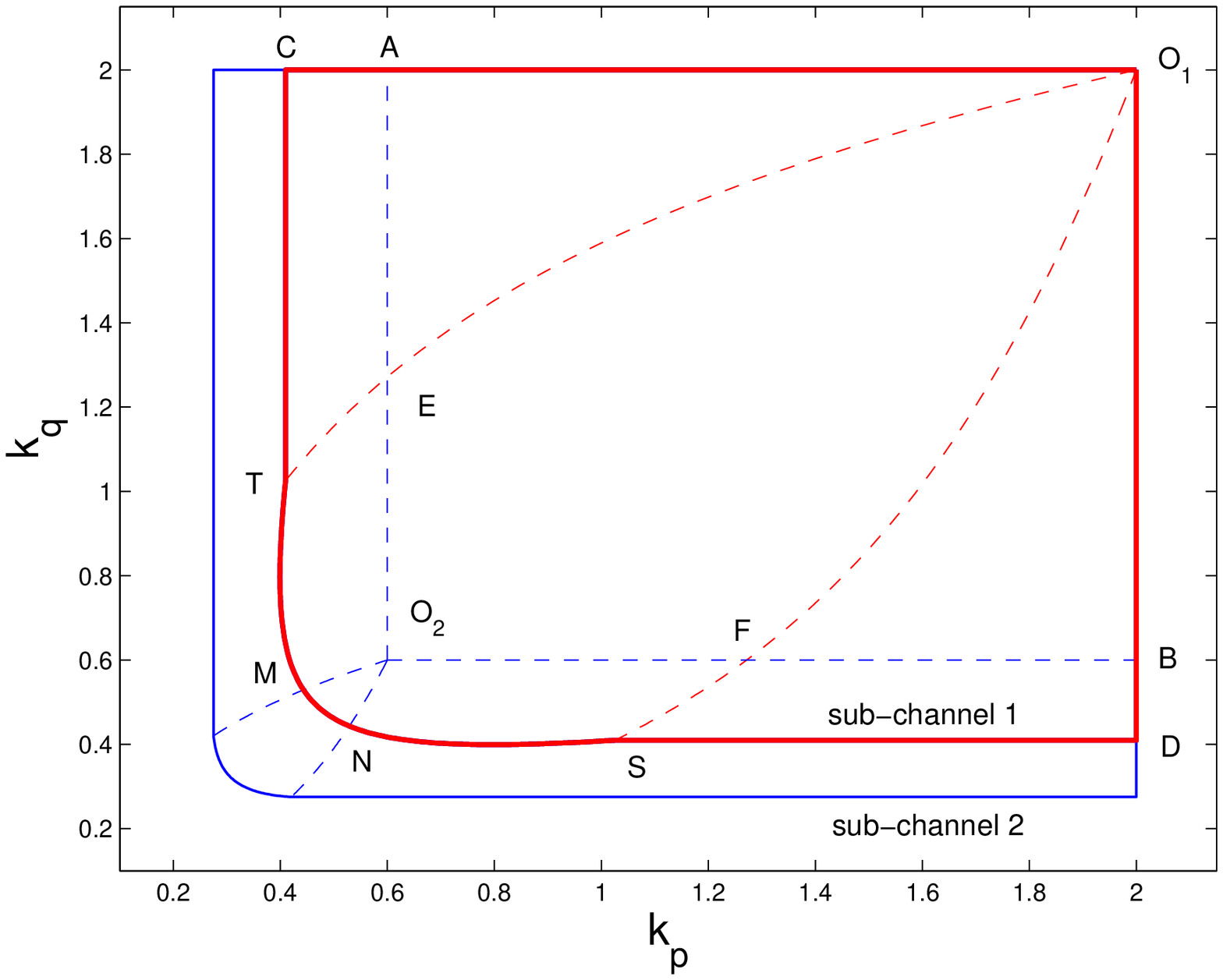}}\caption{$\Bmat_1$ and $\Bmat_2$ for the
parallel Gaussian interference channel with
$a_1=b_1=0.6,c_1=d_1=4,a_2=b_2=0.24,c_2=d_2=1.2$.}
\label{fig:intersect}\end{figure}

\begin{figure}[htp]
\centerline{\leavevmode \epsfxsize=5.5in \epsfysize=4.5in
\epsfbox{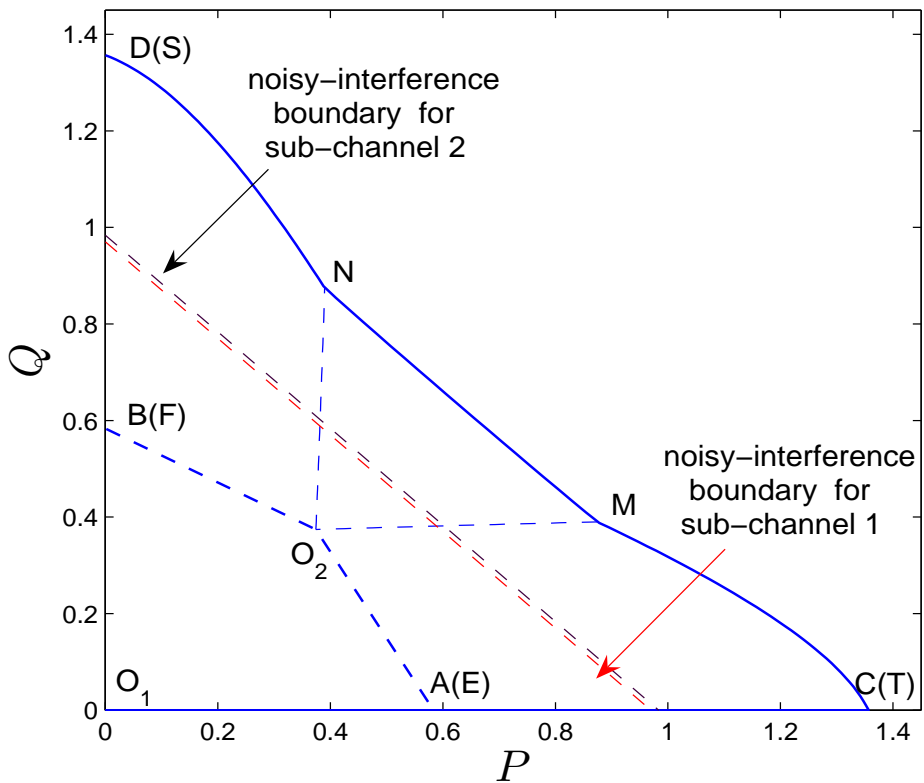}}\caption{Noisy-interference power region
for the parallel Gaussian interference channel with
$a_1=b_1=0.6,c_1=d_1=4,a_2=b_2=0.24,c_2=d_2=1.2$.}
\label{fig:intersectP}\end{figure}

Fig. \ref{fig:intersect} shows the regions of $\Bmat_1$ and
$\Bmat_2$ for a PGIC with two sub-channels. In this case
$\Bmat_1\subset\Bmat_2$. In Fig. \ref{fig:intersectP}, ${O_1CMND}$
is the noisy-interference power region for this PGIC. Points $A,E$
coincide in Fig. \ref{fig:intersectP} since in Fig.
\ref{fig:intersect} $A,E\in \Bmat_1^{(2)}\bigcap\Bmat_2^{(4)}$ and
thus the power allocation is $Q_1^*=P_2^*=Q_2^*=0$ and $P_1^*>0$.
Similarly, points $C,T$ coincide in Fig. \ref{fig:intersectP}
since $C,T\in\Bmat_1^{(2)}\bigcap\Bmat_2^{(2)}$ and thus
$Q_1^*=Q_2^*=0$ and $P_1^*>0,P_2^*>0$. Similar arguments apply to
points $B,F$ and $D,S$. By considering the relationship between
the power regions and the respective subdifferentials we can
determine the activity of the two users in each sub-channel. This
activity is summarized in Tab. \ref{tab:powerRegion}, where $0$
indicates inactive (zero power) and $+$ indicates active (positive
power) for the user in the corresponding sub-channel. In the
following we illustrate the regions in Figs. \ref{fig:intersect}
and \ref{fig:intersectP}.

We first remind the reader that $\Bmat_i^{(1)}$ corresponds to the
case where both users are active; $\Bmat_i^{(2)}$ corresponds to
the case where only user $1$ is active; $\Bmat_i^{(3)}$
corresponds to the case where only user $2$ is active; and
$\Bmat_i^{(4)}$ corresponds to the case where both users are
inactive in sub-channel $i$. Consider the following regions.
\bi%
\item The region $O_2MN$ in Tab. \ref{tab:powerRegion} denotes
regions in both Figs. \ref{fig:intersect} and
\ref{fig:intersectP}. In Fig. \ref{fig:intersect}, it is the
intersection of $\Bmat_1^{(1)}$ and $\Bmat_2^{(1)}$. Thus, to
achieve the sum-rate capacity both users are active in both
sub-channels. The corresponding power region $O_2MN$ is shown in
Fig. \ref{fig:intersectP}.
\item Region $O_1EO_2F$ is the intersection of $\Bmat_1^{(1)}$ and
$\Bmat_2^{(4)}$. So both users are active only in sub-channel $1$.
In this case, both of the power constraints $P$ and $Q$ are small,
so that only the better sub-channel which produces larger sum-rate
capacity than the other is allocated power. Therefore, this
two-channel PGIC behaves as a GIC.
\item Region $O_1AE$ is the intersection of $\Bmat_1^{(2)}$ and
$\Bmat_2^{(4)}$. So user $1$ is active in sub-channel $1$ and user
$2$ is inactive in both sub-channels. The power region $O_1AE$ is
shown in Fig. \ref{fig:intersectP}. The overall power for user $2$
is $Q=0$.
\item Similar to the above case, region $AETC$ of Fig.
\ref{fig:intersect} is the intersection of $\Bmat_1^{(2)}$ and
$\Bmat_2^{(2)}$. The optimal power allocation makes user $1$
active in both sub-channels while user $2$ is inactive in both
sub-channels. In Fig. \ref{fig:intersectP}, the overall power for
user $2$ is also $Q=0$. Actually, regions $O_1AE$ and $AETC$ are
 examples of single-user parallel Gaussian channels whose
optimal power allocation is the water-filling scheme. In the
former case, the power constraint is so small that only the
sub-channel with larger direct link channel gain (sub-channel $1$)
is allocated power. In the latter case, the power constraint
increases to a critical level (point $A$ in Fig.
\ref{fig:intersectP}) so that both sub-channels are allocated
power.
\item Region $ETMO_2$ is the intersection of $\Bmat_1^{(1)}$ and
$\Bmat_2^{(2)}$. User $1$ is active in both sub-channels while
user $2$ is active only in sub-channel $1$ which has larger direct
channel gain for user $2$. In this case, the two-channel PGIC with
two-sided interference works like a PGIC with one sub-channel
having two-sided interference and the other having one-sided
interference.
\item Regions $O_1BF$, $BFSD$ and $FO_2NS$ are counterparts of
regions $O_1AE$, $AETC$ and $ETMO_2$, respectively, by swapping
the roles of the two users. %
\ei

 Also plotted in
Fig. \ref{fig:intersectP} are the noisy-interference power regions
for the individual sub-channels, where sub-channel $2$ has a
larger noisy-interference power region than that of sub-channel
$1$. In this case, the overall noisy-interference power region is
larger than that of either of these two sub-channels.

\begin{table}[htp]\caption{\label{tab:powerRegion} Power constraints and the activeness of
users}\vspace{.1in}
{\centerline{\begin{tabular}{|c|c|c|c|c|c|c|c|c|}
  \hline
   & $O_1A(E)$ & $A(E)(T)C$ & $ETMO_2$ & $O_1EO_2F$  \\
  \hline
    & $\Bmat_1^{(2)}\bigcap\Bmat_2^{(4)}$ & $\Bmat_1^{(2)}\bigcap\Bmat_2^{(2)}$ & $\Bmat_1^{(1)}\bigcap\Bmat_2^{(2)}$ & $\Bmat_1^{(1)}\bigcap\Bmat_2^{(4)}$   \\
    \hline
  $(P_1^*,Q_1^*)$ & (+,0) & (+,0) & (+,+) & (+,+)  \\
  $(P_2^*,Q_2^*)$ & (0,0) & (+,0) & (+,0) & (0,0)  \\
  \hline
    \hline
   & $O_2MN$ & $O_1B(F)$ & $B(F)(S)D$ & $FO_2NS$ \\
  \hline
    &  $\Bmat_1^{(1)}\bigcap\Bmat_2^{(1)}$ & $\Bmat_1^{(3)}\bigcap\Bmat_2^{(4)}$ & $\Bmat_1^{(3)}\bigcap\Bmat_2^{(3)}$ & $\Bmat_1^{(1)}\bigcap\Bmat_2^{(3)}$ \\
    \hline
  $(P_1^*,Q_1^*)$ &  (+,+) & (0,+) & (0,+) & (+,+) \\
  $(P_2^*,Q_2^*)$ &  (+,+) & (0,0) & (0,+) & (0,+) \\
  \hline
\end{tabular}}}
\end{table}

\section{Conclusion} \label{section:conclusion}
Based on the concavity of the sum-rate capacity in the power
constraints, we have shown that the noisy-interference sum-rate
capacity of a PGIC can be achieved by independent transmission
across sub-channels and treating interference as noise in each
sub-channel. The optimal power allocations have the property that
the sub-channel sum-rate capacity curves have parallel supporting
hyperplanes at these powers. The methods introduced in this paper
can also be used to obtain the optimal power allocation and
capacity regions of parallel Gaussian multiple access and
broadcast channels \cite{Shang:thesis}.

\section*{Appendix: subdifferential of $C_i(p,q)$}
\label{section:appendix}%
 The subdifferential $\partial C_i(p,q)$
depends on the location of the point $(p,q)$. If the sub-channel
is a two-sided GIC, from Lemma \ref{lemma:PGIC_NIsumCapacity},
$\Amat_i$ is defined in (\ref{eq:PGIC_NI}). We derive the
subdifferentials as outlined in
(\ref{eq:subdbegin})-(\ref{eq:subdend}) below and present the
evaluations in (\ref{eq:subd}) below.

 \bi \item If $(p,q)$ is an interior point of
$\Amat_i$, $C_i(p,q)$ is differentiable. From the concavity of
$C_i(p,q)$, $\triangledown C_i(p,q)$ satisfies (\ref{eq:subg}) and
the subdifferential $\partial C_i(p,q)$ consists of a unique
vector $[k_p,k_q]^T=\triangledown C_i(p,q)$ where
\bqa%
k_p=\frac{\partial C_i(p,q)}{\partial p}\label{eq:subdbegin}\\
k_q=\frac{\partial C_i(p,q)}{\partial q}%
\eqa

\item If
$\sqrt{a_ic_i}(1+b_ip)+\sqrt{b_id_i}(1+a_iq)=\sqrt{c_id_i},p> 0,q>
0$, we can compute only one-sided partial derivatives since
$C_i(p,q)$ is unknown on the other side. The subdifferential
includes a vector $[k_p,k_q]^T$ where
\bqa%
 k_p=\lim_{\delta\uparrow
0}\frac{C_i(p+\delta,q)-C_i(p,q)}{\delta},\\
k_q=\lim_{\delta\uparrow
0}\frac{C_i(p,q+\delta)-C_i(p,q)}{\delta},%
\eqa
where $\displaystyle\lim_{\delta\uparrow
0}\triangleq\lim_{\delta\leq 0,\delta\rightarrow 0}$ and similarly
$\displaystyle\lim_{\delta\downarrow 0}\triangleq\lim_{\delta\geq
0,\delta\rightarrow 0}$

\item If $q=0,p>0$, we have only a one-sided partial derivative in
$q$. From the concavity of $C_i(p,q)$, all the $[k_p,k_q]^T$
vectors satisfying following conditions are subgradients \bqa
&&k_p=\left\{\begin{array}{ll}
  \dfrac{\partial C_i(p,0)}{\partial p}, & \quad 0<p<\dfrac{\sqrt{c_id_i}-\sqrt{a_ic_i}-\sqrt{b_id_i}}{b_i\sqrt{a_ic_i}}, \\
  \displaystyle\lim_{\delta\uparrow 0}\dfrac{C_i(p+\delta,0)-C_i(p,0)}{\delta},
  & \quad
  p=\dfrac{\sqrt{c_id_i}-\sqrt{a_ic_i}-\sqrt{b_id_i}}{b_i\sqrt{a_ic_i}},
\end{array}\right.\\
&&k_q=\lim_{\delta\downarrow
0}\dfrac{C_i(p,\delta)-C_i(p,0)}{\delta}.\label{eq:kqq0}\eqa On
the other hand, for the same point $(p,q)$ and the corresponding
$[k_p,k_q]^T$ defined above, by choosing
$k_q^\prime>\lim_{\delta\downarrow
0}\dfrac{C_i(p,\delta)-C_i(p,0)}{\delta}$, the vector
$[k_p,k_q]^T$ satisfies (\ref{eq:subg}) for all points
$[p,q]^T\in\Amat_i$, and thus is also a subgradient. Therefore we
can replace (\ref{eq:kqq0}) with \bqa
k_q\geq\lim_{\delta\downarrow
0}\dfrac{C_i(p,\delta)-C_i(p,0)}{\delta}.\eqa

\item Similarly, if $p=0,q>0$, the subdifferential is the set of
$[k_p,k_q]^T$ with
\bqa%
 &&k_p\geq\lim_{\delta\downarrow
0}\dfrac{C_i(\delta,q)-C_i(0,q)}{\delta},\\
&&k_q=\left\{\begin{array}{ll}
  \dfrac{\partial C_i(0,q)}{\partial q}, & \quad 0<q<\dfrac{\sqrt{c_id_i}-\sqrt{a_ic_i}-\sqrt{b_id_i}}{a_i\sqrt{b_id_i}}, \\
  \displaystyle\lim_{\delta\uparrow 0}\dfrac{C_i(p+\delta,0)-C_i(p,0)}{\delta},
  & \quad
  q=\dfrac{\sqrt{c_id_i}-\sqrt{a_ic_i}-\sqrt{b_id_i}}{a_i\sqrt{b_id_i}}.
\end{array}\right.%
\eqa
\item If $p=q=0$ the subdifferential is the set of $[k_p,k_q]^T$
with
\bqa%
 k_p\geq\lim_{\delta\downarrow
0}\frac{C_i(\delta,0)-C_i(0,0)}{\delta},\\
k_q\geq\lim_{\delta\downarrow
0}\frac{C_i(0,\delta)-C_i(0,0)}{\delta}.%
\label{eq:subdend}%
\eqa
 \ei

For completeness, we summarize $\partial C_i(p,q)$ as follows \bqa
&&\partial C_i(p,q)\nn\\
&&=\left\{\begin{array}{ll}
  \left\{(k_p,k_q)\left|\begin{array}{l}
  k_p=\dfrac{1}{2}\left(\dfrac{c_i}{1+c_ip+a_iq}+\dfrac{b_i}{1+b_ip+d_iq}-\dfrac{b_i}{1+b_ip}\right) \\
  k_q=\dfrac{1}{2}\left(\dfrac{a_i}{1+c_ip+a_iq}+\dfrac{d_i}{1+b_ip+d_iq}-\dfrac{a_i}{1+a_iq}\right) \\
\end{array}\right.\right\}, & \quad (p,q)\in\Amat_i^{(1)} \\
  \left\{(k_p,k_q)\left|\begin{array}{l}
  k_p=\dfrac{c_i}{2(1+c_ip)} \\
  \dfrac{d_i}{2(1+b_ip)}-\dfrac{a_ic_ip}{2(1+c_ip)}\leq k_q\leq\dfrac{\hat d}{2} \\
\end{array}\right.\right\}, & \quad (p,q)\in\Amat_i^{(2)} \\
  \left\{(k_p,k_q)\left|\begin{array}{l}
  \dfrac{c_i}{2(1+a_iq)}-\dfrac{b_id_iq}{2(1+d_iq)}\leq k_p\leq\dfrac{\hat c}{2} \\
  k_q=\dfrac{d_i}{2(1+d_iq)} \\
\end{array}\right.\right\}, & \quad (p,q)\in\Amat_i^{(3)} \\
  \left\{(k_p,k_q)\left|\dfrac{c_i}{2}\leq
k_p\leq\dfrac{\hat c}{2} ,\quad\dfrac{d_i}{2}\leq k_q\leq
\dfrac{\hat
d}{2}\right.\right\}, & \quad (p,q)\in\Amat_i^{(4)}. \\
\end{array}\right.\label{eq:subd}\eqa
where \bqa
&&\Amat_i^{(1)}=\left\{(p,q)\left|\sqrt{a_i}(1+b_ip)+\sqrt{b_i}(1+a_iq)\leq\sqrt{c_id_i},p>
0,q> 0\right.\right\},\\
&&\Amat_i^{(2)}=\left\{(p,q)\left|0<p\leq\frac{\sqrt{c_id_i}-\sqrt{a_ic_i}-\sqrt{b_id_i}}{b_i\sqrt{a_ic_i}},q=0\right.\right\},\\
&&\Amat_i^{(3)}=\left\{(p,q)\left|p=0,0<q\leq\frac{\sqrt{c_id_i}-\sqrt{a_ic_i}-\sqrt{b_id_i}}{a_i\sqrt{b_id_i}}\right.\right\},\\
&&\Amat_i^{(4)}=\left\{(p,q)\left|p=q=0\right.\right\},\eqa and
\bqa &&\hat c=\max_{i=1,\cdots,m}\{c_i\},\\
&&\hat d=\max_{i=1,\cdots,m}\{d_i\}.\eqa In (\ref{eq:subd})
$\Amat_i=\bigcup_{l=1}^4\Amat_i^{(l)}$. For the subgradient
$[k_p,k_q]^T$, when $p=0$ or $q=0$, $k_p$ or $k_q$ varies from
some constants to infinity.  In (\ref{eq:subd}) we upper bound
$k_p$ and $k_q$ with $\frac{\hat c}{2}$ and $\frac{\hat d}{2}$
respectively for convenience and without loss of generality. The
main reason is that we are interested only in
$\bigcup_{i=1}^m\Bmat_i$. To relate the mapping of $\Amat_i^{(j)}$
to different regions of $\Bmat_i$, we further define
\bqa%
\Bmat_i^{(j)}=\bigcup_{[p,q]^T\in\Amat_i^{(j)}}\partial C_i(p,q),\quad j=1,\cdots,4. %
\eqa

If the sub-channel has one-sided interference $b_i=0,0<a_i<1$ or
no interference $a_i=b_i=0$, then from Lemma
\ref{lemma:PGIC_NIsumCapacity} we have $\Amat_i=\{p\geq 0,q\geq
0\}$. We similarly obtain $\partial C_i(p,q)$ as follows:
\bqa%
&&\partial C_i(p,q)\nn\\
&&=\left\{\begin{array}{ll}
  \left\{(k_p,k_q)\left|\begin{array}{l}
  k_p=\dfrac{c_i}{2(1+c_ip+a_iq)} \\
  k_q=\dfrac{1}{2}\left(\dfrac{a_i}{1+c_ip+a_iq}+\dfrac{d_i}{1+d_iq}-\dfrac{a_i}{1+a_iq}\right) \\
\end{array}\right.\right\}, &\quad p>0,q>0 \\
  \left\{(k_p,k_q)\left|\begin{array}{l}
  k_p=\dfrac{c_i}{2(1+c_ip)} \\
  \dfrac{d_i}{2}-\dfrac{a_ic_ip}{2(1+c_ip)}\leq k_q\leq\dfrac{\hat d}{2} \\
\end{array}\right.\right\}, & \quad p>0,q=0 \\
  \left\{(k_p,k_q)\left|\begin{array}{l}
  \dfrac{c_i}{2(1+a_iq)}\leq k_p\leq\dfrac{\hat c}{2} \\
  k_q=\dfrac{d_i}{2(1+d_iq)} \\
\end{array}\right.\right\}, & \quad p=0,q>0 \\
  \left\{(k_p,k_q)\left|\dfrac{c_i}{2}\leq
k_p\leq\dfrac{\hat c}{2} ,\quad\dfrac{d_i}{2}\leq k_q\leq
\dfrac{\hat
d}{2}\right.\right\}, & \quad p=q=0. \\
\end{array}\right.%
\label{eq:ZICsubd}%
\eqa
\bibliography{Journal,Conf,Misc,Book}

\begin{thebibliography}{10}

\bibitem{Chung&Cioffi:07COM}
S.~T. Chung and J.~M. Cioffi,
\newblock ``{The capacity region of frequency-selective Gaussian interference
  channels under strong interference},''
\newblock {\em IEEE Trans. Communications}, vol. 55, pp. 1812--1821, Sept.
  2007.

\bibitem{Yu-etal:02JSAC}
W.~Yu, G.~Ginis, and J.~M. Cioffi,
\newblock ``{Distributed multiuser power control for digital subscriber
  lines},''
\newblock {\em IEEE Journal on Selected Areas in Communications}, vol. 20, pp.
  1105--1115, 2002.

\bibitem{Song-etal:02COMMAG}
K.~B. Song, S.~T. Chung, G.~Ginis, and J.~M. Cioffi,
\newblock ``{Dynamic spectrum management for next-generation DSL systems},''
\newblock {\em IEEE Communications Magazine}, vol. 40, pp. 101--109, 2002.

\bibitem{Cendrillon-etal:06COM}
R.~Cendrillon, W.~Yu, M.~Moonen, J.~Verlinden, and T.~Bostoen,
\newblock ``{Optimal multi-user spectrum management for digital subscriber
  lines},''
\newblock {\em IEEE Trans. Communications}, vol. 54, pp. 922--933, 2006.

\bibitem{Hayashi&Luo:ITsubmission}
S.~Hayashi and Z.-Q. Luo,
\newblock ``{Spectrum management for interference-limited multiuser
  communication systems},''
\newblock {\em {\em submitted to} IEEE Trans. Inf. Theory}.

\bibitem{Shang-etal:09IT}
X.~Shang, G.~Kramer, and B.~Chen,
\newblock ``{A new outer bound and the noisy-interference sum-rate capacity for
  Gaussian interference channels},''
\newblock {\em IEEE Trans. Inf. Theory}, vol. 55, no. 2, pp. 689--699, Feb.
  2009.

\bibitem{Motahari&Khandani:08IT_submission}
A.~S. Motahari and A.~K. Khandani,
\newblock ``{Capacity bounds for the Gaussian interference channel},''
\newblock {\em {\em submitted to} IEEE Trans. Inf. Theory.
  http://arxiv.org/abs/0801.1306}, Jan. 2008.

\bibitem{Annapureddy&Veeravalli:08IT_submission}
V.~S. Annapureddy and V.~Veeravalli,
\newblock ``{Gaussian interference networks: Sum capacity in the low
  interference regime and new outer bounds on the capacity region},''
\newblock {\em {\em submitted to} IEEE Trans. Inf. Theory.
  http://arxiv.org/abs/0802.3495}, Feb. 2008.

\bibitem{Shang-etal:08Allerton}
X.~Shang, B.~Chen, G.~Kramer, and H.~V. Poor,
\newblock ``{On the capaicty of MIMO interference channels},''
\newblock in {\em Proc. of the Forty-sixth Annual Allerton Conference on
  Communication, Control, and Computing}, Monticello, IL, Sep. 2008.

\bibitem{Tse&Hanly:98IT}
D.~N.~C. Tse and S.~V. Hanly,
\newblock ``{Multiaccess fading channels¨Cpart I: polymatroid structure,
  optimal resource allocation and throughput capacities},''
\newblock {\em IEEE Trans. Inf. Theory}, vol. 44, no. 7, pp. 2796--2815, Nov.
  1998.

\bibitem{Elgamel:80Problemy}
A.~El Gamal,
\newblock ``{The capacity of the product and sum of two unmatched degraded
  broadcast channels},''
\newblock {\em Problemy Perdachi Informatsi}, vol. 6, no. 1, pp. 3--23, 1980.

\bibitem{Hughes-Hartog:thesis}
D.~Hughes-Hartog,
\newblock {\em The Capacity of a Degraded Spectral Gaussian Broadcast
  Channel,},
\newblock Ph.D. thesis, Stanford University, Stanford, CA, Jul. 1995.

\bibitem{Tse:98net}
D.~N.~C. Tse,
\newblock ``{Optimal power allocation over parallel Gaussian broadcast
  channels},''
\newblock {\em available at
  http://www.eecs.berkeley.edu/$\sim$dtse/broadcast2.pdf}, 1998.

\bibitem{Cadambe&Jafar:08IT_submission}
S.~A.~Jafar V.~R.~Cadambe,
\newblock ``{Multiple access outerbounds and the inseparability of parallel
  interference channels},''
\newblock {\em {\em submitted to }IEEE Trans. Inf. Theory.
  http://arxiv.org/abs/0802.2125}, Feb. 2008.

\bibitem{Sankar-etal:08Allerton}
L.~Sankar, X.~Shang, E.~Erkip, and H.~V. Poor,
\newblock ``{Ergodic two-user interference channels: is separability
  optimal},''
\newblock in {\em Proceedings of the Forty-sixth Annual Allerton Conference on
  Communication, Control, and Computing}, Monticello, IL, Sep. 2008.

\bibitem{Carleial:75IT}
A.~B. Carleial,
\newblock ``{A case where interference does not reduce capacity},''
\newblock {\em IEEE Trans. Inf. Theory}, vol. 21, pp. 569--570, Sep. 1975.

\bibitem{Sato:81IT}
H.~Sato,
\newblock ``{The capacity of the Gaussian interference channel under strong
  interference},''
\newblock {\em IEEE Trans. Inf. Theory}, vol. 27, pp. 786--788, Nov. 1981.

\bibitem{Han&Kobayashi:81IT}
T.~S. Han and K.~Kobayashi,
\newblock ``{A new achievable rate region for the interference channel},''
\newblock {\em IEEE Trans. Inf. Theory}, vol. 27, pp. 49--60, Jan. 1981.

\bibitem{Sato:78IT}
H.~Sato,
\newblock ``{On degraded Gaussian two-user channels},''
\newblock {\em IEEE Trans. Inf. Theory}, vol. 24, pp. 634--640, Sep. 1978.

\bibitem{Bertsekas:03book}
D.~P. Bertsekas,
\newblock {\em Nonlinear Programming},
\newblock Athena Scientific, Belmont, MA, 2003.

\bibitem{Thomas:87IT}
J.~A. Thomas,
\newblock ``{Feedback can at most double Gaussian multiple access channel
  capacity},''
\newblock {\em IEEE Trans. Inf. Theory}, vol. 33, no. 5, pp. 711--716, Sep.
  1987.

\bibitem{Shang:thesis}
X.~Shang,
\newblock {\em On the Capacity of Gaussian Interference Channels},
\newblock Ph.D. thesis, Syracuse University, Syracuse, NY, Aug. 2008.

\end{thebibliography}
\bibliographystyle{IEEEbib}
\end{document}